\documentclass[journal]{IEEEtran}
\usepackage{amsmath,amsfonts}
\usepackage{algorithmic}
\usepackage{algorithm}
\usepackage{array}
\usepackage[caption=false,font=normalsize,labelfont=sf,textfont=sf]{subfig}
\usepackage{textcomp}
\usepackage{stfloats}
\usepackage{url}
\usepackage{verbatim}
\usepackage{graphicx}
\usepackage{cite}
\usepackage{multirow}
\usepackage{booktabs}
\usepackage{soul}
\usepackage[normalem]{ulem}

\newcolumntype{L}{>{\RaggedRight\hangafter=1\hangindent=0em}X}
\usepackage{tikz,xcolor,hyperref}
\hypersetup{hidelinks}
\definecolor{lime}{HTML}{A6CE39}
\DeclareRobustCommand{\orcidicon}{%
	\begin{tikzpicture}
	\draw[lime, fill=lime] (0,0) 
	circle [radius=0.16] 
	node[white] {{\fontfamily{qag}\selectfont \tiny ID}};
	\draw[white, fill=white] (-0.0625,0.095) 
	circle [radius=0.007];
	\end{tikzpicture}
	\hspace{-2mm}
}

\foreach \x in {A, ..., Z}{%
	\expandafter\xdef\csname orcid\x\endcsname{\noexpand\href{https://orcid.org/\csname orcidauthor\x\endcsname}{\noexpand\orcidicon}}
}

\begin{document}

\title{BECS: A Privacy-Preserving Computing Resource Sharing Mechanism for 6G Computing Power Network}

\author{Kun Yan\orcidA{},~\IEEEmembership{Member,~IEEE}, 
        Wenping Ma\orcidB{},~\IEEEmembership{Member,~IEEE}, 
		Shaohui Sun\orcidC{},~\IEEEmembership{Senior Member,~IEEE}


\thanks{\textcolor{red}{This paper has been accepted by IEEE Transactions on Network Science and Engineering, and the copyright belongs to IEEE.} This work was supported in part by the Open Fund of the State Key Laboratory of Intelligent Vehicle Safety Technology, under Grant IVSTSKL-202441. \emph{(Corresponding author: Wenping Ma.)}}
\thanks{Kun Yan and Wenping Ma are with the School of Telecommunications Engineering, Xidian University, Xi'an 710071, China, and also with the State Key Laboratory of Intelligent Vehicle Safety Technology, Changan Automobile Company, Ltd., Chongqing 400044, China (e-mail: kyan@stu.xidian.edu.cn; wp\_ma@mail.xidian.edu.cn).}
\thanks{Shaohui Sun is with the State Key Laboratory of Wireless Mobile Communication, Datang Mobile Communications Equipment Co., Ltd., Beijing 100083, China (e-mail: sunshaohui@catt.cn).}
}

\markboth{IEEE Transactions on Network Science and Engineering,~Vol.~14, No.~8, April~2025}%
{Shell \MakeLowercase{\textit{et al.}}: A Sample Article Using IEEEtran.cls for IEEE Journals}

\maketitle

\begin{abstract}

The 6G Computing Power Network (CPN) is envisioned to orchestrate vast, distributed computing resources for future intelligent applications. However, achieving efficient, trusted, and privacy-preserving computing resource sharing in this decentralized environment poses significant challenges. To address these intertwined issues, this article proposes a holistic blockchain and evolutionary algorithm-based computing resource sharing (BECS) mechanism. BECS is designed to dynamically and adaptively balance task offloading among computing resources within the 6G CPN, thereby enhancing resource utilization. We model computing resource sharing as a multi-objective optimization problem, aiming to navigate these trade-offs. To tackle this NP-hard problem, we devise a kernel-distance-based dominance relation and incorporate it into the Non-dominated Sorting Genetic Algorithm III (NSGA-III), thereby significantly enhancing population diversity. In addition, we propose a pseudonym scheme based on zero-knowledge proofs to protect user privacy during computing resource sharing. Finally, security analysis and simulation results demonstrate that BECS can effectively leverage all computing resources in the 6G CPN, thereby significantly improving resource utilization while preserving user privacy.

\end{abstract}

\begin{IEEEkeywords}
6G computing power network, computing resource sharing, multi-objective evolutionary optimization, blockchain, pseudonym scheme.
\end{IEEEkeywords}

\section{Introduction} 
\IEEEPARstart{I}{n} the upcoming 6G era, it is anticipated that everything will be intelligently connected, enabling a wide range of data-intensive applications. By deeply integrating communication networks with various vertical industries, 6G will enable unprecedented applications such as Holographic Integrated Sensing and Communication (HISC), Artificial General Intelligence (AGI), and Digital Twins (DT) \cite{1,65,66}. This indicates that AI will be one of the most crucial technologies for constructing a comprehensively intelligent 6G networks, enabling network services to evolve dynamically and autonomously in response to demand. The coordination of end-edge-cloud computing devices to form a Computing Power Network (CPN) is expected to become a leading paradigm for supporting ubiquitous intelligent services in 6G networks \cite{2}. In essence, the CPN envisions intelligent orchestration, where computing tasks are dynamically offloaded to the most suitable execution environment based on real-time service demands. This fosters a highly dynamic and hybrid computing environment that promotes the sharing of computing resources across 6G networks. However, realizing this vision in the 6G CPN presents fundamental challenges. In particular, it is necessary to design a resource sharing mechanism that can simultaneously ensure trustworthiness and preserve user privacy, while navigating the complex performance trade-offs inherent in such a massive and decentralized system.

While Multi-access Edge Computing (MEC) and fog computing in 5G networks paved the way for offloading computing tasks to the network edge and unleashing potential for the Internet of Things (IoT) \cite{47}, their limitations become evident when faced with more complex demands in the future. These paradigms are typically characterized by localized deployments, hierarchical interactions, and siloed management, all of which hinder the seamless integration and orchestration of ubiquitously distributed, heterogeneous computing resources across the network. In the 6G era, networks are expected to scale up to support trillions of connections. Meanwhile, intelligent and diversified applications will not only require computing resources that far exceed previous levels but will also urgently demand a ubiquitous computing paradigm. This paradigm must enable real-time coordination of network-wide computing resources and dynamic on-demand orchestration. Crucially, it involves navigating complex trade-offs among multiple, often conflicting, performance objectives such as minimizing latency, reducing energy consumption, and maximizing resource utilization across seamless cross-domain allocations \cite{48}. The high-dimensional and strongly-coupled nature of these objectives presents a formidable optimization challenge, pushing standard multi-objective algorithms towards their performance limits, particularly in maintaining the solution diversity required to map the entire Pareto-optimal front. As a result, 5G MEC and fog computing are inadequate for meeting future computing demands in terms of both scale and operational paradigms. Therefore, 6G networks require the CPN to deeply integrate and intelligently orchestrate dispersed, heterogeneous, and geographically distributed computing resources across the entire network. Such integration promotes full resource utilization and enables efficient, large-scale computing resource sharing through collaborative offloading and incentive mechanisms \cite{67,68,69}. 

Driven by the unprecedented features and services of future 6G networks, communication networks are increasingly exploring distributed or multi-center management models \cite{49}. Furthermore, the 6G CPN aims to comprehensively mobilize vast heterogeneous devices, edge nodes, and cloud computing resources across organizational boundaries. This cross-domain and heterogeneous environment, which lacks a central authority of trust, poses significant challenges for access, scheduling, trading, and management of computing resources \cite{50}. Traditional centralized architectures not only struggle with inefficiency and single points of failure, but they also lack the transparency and immutability needed to build trust among diverse stakeholders. Therefore, it is necessary to establish a novel distributed management mechanism that enables autonomous, reliable, and efficient collaboration among devices. Such a mechanism must provide verifiable, tamper-proof records for resource trading and enforce operational rules transparently, without relying on a central intermediary. Blockchain, with its inherent decentralization and its ability to enhance trust and transparency among multiple stakeholders, emerges as a key enabling technology for future 6G network management \cite{51,70,71}. It offers secure, transparent, and traceable solutions for computing resource sharing, trusted trading, access control, and other critical functionalities within the CPN.

Furthermore, the ubiquitous sharing and collaboration of computing resources exacerbate the risk of user privacy leakage. Distinct from the localized hierarchical structure of fog computing, the 6G CPN orchestrates a wide variety of computing devices into a unified, network-wide computing continuum. In particular, when user devices participate in computing resource sharing, not only their identity but also their behavioral patterns and interaction records can be collected and linked, allowing attackers to infer sensitive information \cite{52,86,87}. These potential threats create a significant barrier to participation. To achieve efficient computing resource sharing, it is essential to provide a mechanism that allows users to prove their legitimacy and conduct transactions without revealing their persistent identities, thus ensuring both accountability and anonymity.

To address these intertwined challenges of efficiency, trust, and privacy, this article introduces a novel mechanism for 6G CPN, named BECS, which utilizes \textbf{B}lockchain and \textbf{E}volutionary algorithms for efficient \textbf{C}omputing \textbf{S}haring. It is designed to create a robust and efficient ecosystem where heterogeneous computing resources can be shared on a large scale. By effectively tackling the critical issues of resource allocation efficiency, decentralized trust, and user privacy preservation, BECS aims to unlock the full potential of distributed computing in 6G networks. Performance evaluations confirm that the proposed BECS mechanism improves average resource utilization by 56.94\% compared to an NSGA-III-based mechanism. Specifically, the main contributions of this article are summarized as follows:

\begin{itemize}
\item{We propose a dynamic and efficient blockchain-based mechanism for computing resource sharing, aimed at ensuring secure allocation and trading of resources between any devices in 6G CPN, thus enhancing utilization.}
\item{We formulate computing allocation as a comprehensive multi-objective optimization problem (MOOP) with six objectives, employing an evolutionary algorithm to balance the interplay among these objectives, thus achieving optimal allocation schemes for 6G CPN.}
\item{We propose a novel evolutionary algorithm, NSGA-III-KDR, which improves the dominance relation of Non-dominated Sorting Genetic Algorithm III (NSGA-III) by using the kernel distance to enhance diversity in addressing the computing allocation MOOP.}
\item{We design a novel pseudonym scheme based on the Schnorr protocol, which protects user privacy during computing resource sharing in 6G CPN.}
\end{itemize}

The remainder of this article is organized as follows. Section II reviews related work. Section III introduces the system model and formulates the computing resource sharing problem. Section IV describes NSGA-III-KDR and its application in solving the computing allocation MOOP. Section V presents the proposed pseudonym scheme and computing trading. Section VI analyzes the security and computational complexity of the proposed scheme. Section VII presents simulation results. Section VIII concludes the article.

\section{Related Works}

In this section, we first introduce the paradigm of the 6G CPN and contrast it with 5G MEC. Then, we discuss existing works on computing allocation and sharing. Finally, we elaborate blockchain for decentralized management and privacy preservation.

\subsection{The Paradigm of 6G Computing Power Network}

The trajectory from 5G to 6G signals a profound transformation, moving beyond the traditional pursuit of enhanced communication metrics. Driven by the rapid advancement of AI, it represents a fundamental paradigm shift toward the deep integration of computing and networking \cite{72,73,74}. Although 5G MEC optimizes computing services by pushing computing power to the network edge, its inherent limitations in scale and cross-domain orchestration prevent it from meeting the ubiquitous intelligence demands of 6G \cite{55}.


6G CPN will deeply integrate all computing devices within the network to form a distributed and intelligent computing environment \cite{12}. By seamlessly abstracting, virtualizing, and integrating decentralized computing resources managed by different operators and originating from diverse device types, 6G CPN aims to unify the heterogeneous resource landscape. It constructs a unified computing architecture that spans from user devices to multi-domain edge nodes and cloud centers, capable of supporting complex service architectures \cite{57}. This transition from isolated resources to a unified architecture also assigns different identity roles to computing devices within the network. In 5G MEC or fog computing, devices often operate within fixed hierarchies where user devices are typically resource requesters, while edge/fog nodes act as resource providers. In contrast, 6G CPN breaks this limitation and enables a flat, collaborative ecosystem where every computing device can function both as a requester and a provider, thereby enabling full utilization of all computing resources in the network \cite{58}.

\subsection{Computing Allocation/Sharing in CPN/End-Edge-Cloud}

The dynamic allocation and efficient utilization of computing resources are central to the concept of CPN. Addressing the inherent optimization challenges in managing these aspects has been a key research focus. Lu \emph{et al.} \cite{2} utilized deep reinforcement learning to find the optimal task transfer and computing allocation strategies in wireless CPN. Chen \emph{et al.} \cite{3} proposed an on-demand two-stage computing resource scheduling model to achieve efficient task offloading in edge CPN. However, as 6G services become increasingly diverse and demanding, optimizing for single or dual objectives often proves insufficient to capture the inherent trade-offs between factors like latency, energy consumption, cost, and resource utilization \cite{8}.

\begin{table*}[t]
	\begin{center}
	\caption{Differences between BECS and other main related works}
	\label{diff}
	\renewcommand{\arraystretch}{1.3} 
	\begin{tabular}{|c|>{\centering\arraybackslash}m{0.19\textwidth}|>{\centering\arraybackslash}m{0.18\textwidth}|>
	{\centering\arraybackslash}m{0.05\textwidth}|>{\centering\arraybackslash}m{0.1\textwidth}|>{\centering\arraybackslash}m{0.07\textwidth}|>{\centering\arraybackslash}m{0.06\textwidth}|>{\centering\arraybackslash}m{0.07\textwidth}|}
	\hline
	\textbf{Ref.} & \textbf{Core Scenario} & \textbf{Optimization Method} & \textbf{MOOP} & \textbf{Cloud-Edge-End Collaboration} & \textbf{Blockchain Features} & \textbf{Advanced Privacy Scheme} & \textbf{6G Relevance} \\
	\hline
	\cite{2} & Energy-efficient task transfer in Wireless CPN & Multi-Agent Deep Reinforcement Learning & \texttimes & \texttimes & \texttimes & \texttimes & \checkmark \\
	\hline
	\cite{3} & Efficient task offloading in Edge CPN & Two-Stage Evolutionary Search & \texttimes & \texttimes & \texttimes & \texttimes & \checkmark \\
	\hline
	\cite{6} & Computation offloading in Industrial IoT & NSGA-III & \checkmark & \checkmark & \texttimes & \texttimes & \texttimes \\
	\hline
	\cite{7} & Dependent task offloading in MEC & MOEA/D & \checkmark & \checkmark & \texttimes & \texttimes & \texttimes \\
	\hline
	\cite{53} & Secure computation offloading in IoT & Deep Reinforcement Learning & \texttimes & \checkmark & \checkmark & \texttimes & \texttimes \\
	\hline
    \cite{14} & Cooperative task offloading in MEC & Multi-Agent DRL, Game Theory & \texttimes & \texttimes & \checkmark & \texttimes & \texttimes \\
	\hline
	\cite{15} & Secure computation offloading in cyber-physical systems & Deep Reinforcement Learning & \texttimes & \checkmark & \checkmark & \texttimes & \texttimes \\
	\hline
	\cite{16} & Secure task offloading in MEC & Distributed Deep Q-Learning & \texttimes & \checkmark & \checkmark & \texttimes & \texttimes \\
	\hline
	\textbf{BECS} & Privacy-preserving computing resource sharing in 6G CPN & NSGA-III-KDR & \checkmark & \checkmark & \checkmark & \checkmark & \checkmark \\
	\hline
	\end{tabular}
\end{center}
\end{table*}

Therefore, formulating computing allocation as a MOOP has become a more effective approach for handling these conflicting objectives. Meanwhile, evolutionary algorithms are widely adopted due to their ability to identify a set of non-dominated solutions representing different trade-offs. Peng \emph{et al.} \cite{6} formulated complex task offloading in the IIoT as a four-objective MOOP and developed a method to dynamically allocate computing resources based on the NSGA-III. Gong \emph{et al.} \cite{7} employed the multi-objective evolutionary algorithm based on decomposition (MOEA/D) to optimize a three-objective edge task offloading problem, aiming to minimize delay and maximize rewards. While these approaches demonstrate the applicability of MOEAs, standard algorithms like NSGA-III and MOEA/D face significant challenges when dealing with many-objective optimization, often struggling to maintain sufficient population diversity alongside convergence pressure \cite{27}. This limitation becomes particularly pronounced in the context of our 6G CPN model, where the six-objective formulation and inherent resource heterogeneity necessitate a considerably richer and more diverse set of trade-off solutions than those typically yielded by standard algorithms. Consequently, although these existing tools are capable of generating feasible solutions, their ability to explore the full spectrum of high-quality and well-distributed trade-off strategies remains limited in such a demanding environment.

\subsection{Blockchain for Decentralized Management and Privacy Preservation}

Building on the need for decentralized trust and management in 6G networks, blockchain technology has been actively explored as a key enabler \cite{9,75,80}. Xie \emph{et al.} \cite{10} exploited the immutability of blockchain to propose a resource trading mechanism based on sharding and directed acyclic graphs for large-scale 6G networks, thereby enhancing resource utilization efficiency. Nguyen \emph{et al.} \cite{53} proposed a blockchain-based mobile edge-cloud computation offloading scheme, leveraging the distributed characteristics of blockchain to provide secure and trusted computing services. Wang \emph{et al.} \cite{12} proposed a provable secure blockchain-based federated learning framework for wireless CPN, aimed at accelerating the convergence of federated learning and enhancing the efficiency of wireless CPN. These works highlight blockchain's potential to automate, secure, and streamline interactions like resource discovery, access control, scheduling coordination, and payment settlement via smart contracts \cite{54}, thereby fostering a reliable environment for large-scale computing resource sharing.

6G networks will integrate AI to fully merge the physical and digital worlds, necessitating enhanced security and privacy for robust human-to-virtual connectivity \cite{13,76,77}. Nguyen \emph{et al.} \cite{14} utilized blockchain to provide adequate security for task offloading in mobile edge computing. Wang \emph{et al.} \cite{15} proposed a blockchain-enabled cyber-physical system integrating cloud and edge computing to achieve secure computing offloading. Samy \emph{et al.} \cite{16} introduced a blockchain-based framework for task offloading, ensuring security, integrity, and privacy in mobile edge computing. Although leveraging the characteristics of blockchain can provide preliminary security and privacy protection for edge and cloud computing allocation and trading, device-to-device computing resource sharing in 6G networks will require more comprehensive solutions to ensure the security and privacy between devices \cite{17}.

An overview of related works is given in Table \ref{diff}. Distinct from the aforementioned works, BECS is designed as a holistic mechanism for the 6G CPN. It uniquely combines a blockchain architecture, advanced multi-objective optimization techniques, and cryptographic privacy protection to tackle the intertwined challenges of efficiency and security in computing resource sharing of 6G CPN.

\section{System Model and Problem Formulation}

In this section, we introduce the computing resource sharing system model considered by BECS, along with other models used in constructing the MOOP, including the communication, computing, and service models.

\begin{figure*}[ht]
	\centering
	\includegraphics[scale=0.6]{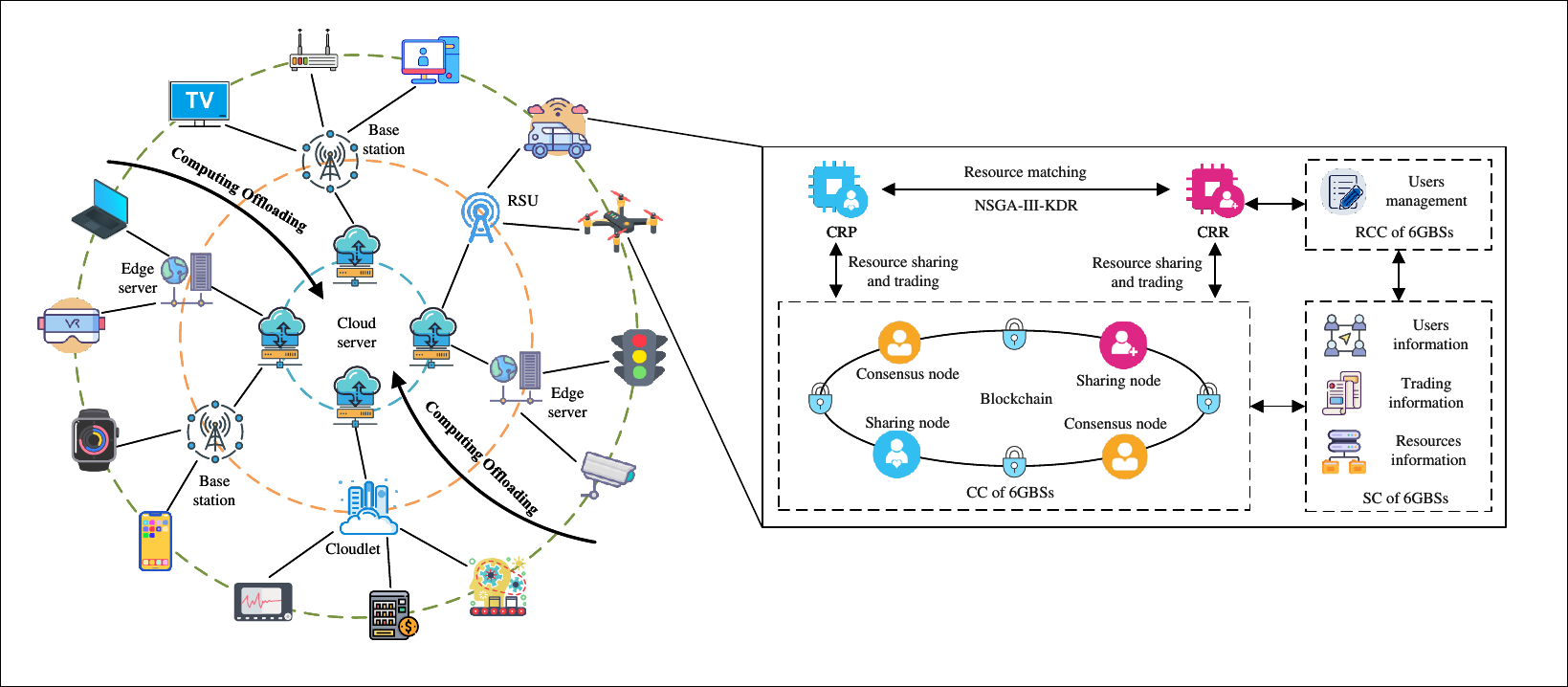}
	\caption{Overview of the BECS architecture in 6G CPN.}
	\label{str}
\end{figure*}

\subsection{System Model}





To meet the demands of future intelligent applications, BECS introduces a generalized architecture, as illustrated in Fig. \ref{str}, designed to support diverse forms of computing resource sharing in 6G CPN. In this architecture, all computing devices, denoted as $\mathbb{D} = \{D_1, \ldots, D_d, \ldots, D_D\}$, are categorized into three layers: the user computing layer, edge computing layer, and cloud computing layer. This classification fundamentally redefines their interplay. Moreover, BECS overcomes the rigid and siloed structure of traditional 5G MEC. Instead of treating computing resources as isolated within discrete physical layers, BECS abstracts and virtualizes all participating devices across the network into an integrated, unified, and orchestrated resource pool. Consequently, the layer classification reflects the functional capabilities of devices rather than enforcing rigid operational boundaries. The resources in $\mathbb{D}$ are inherently heterogeneous, encompassing devices from different administrative domains, multiple operators, and diverse ownerships, thereby forming a cohesive, network-wide computing continuum that can be dynamically orchestrated.

The proposed BECS is primarily designed for a metropolitan area network scenario. In this context, numerous computing devices are geographically distributed across the city and interconnected through a high-bandwidth, low-latency optical fiber backbone. This dense and high-speed infrastructure constitutes a key feature of future 6G-enabled smart city deployments \cite{60}. Crucially, this architecture implies that the main communication bottlenecks and the most significant latency variations primarily arise from wireless access links and task processing, rather than from the highly stable, low-latency wired backbone. Accordingly, the BECS model is strategically designed to optimize these dominant factors. Specifically, the set of user computing devices, denoted as $\mathbb{U} = \{U_1, \ldots, U_u, \ldots, U_U\}$, includes smartphones, computers, wearable devices, IoT devices, vehicles, and other devices that directly interact with users. These devices typically possess limited computing power. The set of edge computing devices, denoted as $\mathbb{E} = \{E_1, \ldots, E_e, \ldots, E_E\}$, includes edge servers, roadside units, cloudlets, and other devices capable of providing time-sensitive computing services to users. The set of cloud computing devices, denoted as $\mathbb{C} = \{C_1, \ldots, C_c, \ldots, C_C\}$, consists of remote computing centers capable of providing large-scale computing power. The evolutionary algorithm matches computing resource requesters with providers, facilitating a multi-dimensional measurement of the deep reuse of computing resources. Additionally, with the support of permissioned blockchain, BECS enables dynamic management and trading of fine-grained computing resources. The proposed architecture consists of three main components:

1) Computing Resource Providers (CRPs): As large language models become increasingly widespread, devices with abundant computing power will increasingly provide computing support to devices with limited resources. In BECS, all devices within 6G CPN with free computing resources can serve as CRPs.

2) Computing Resource Requesters (CRRs): In general, when a device lacks sufficient computing capability to handle a task's demands, it needs to request additional resources. In BECS, any device can be a CRR, provided that the requested resources exceed its own computing capacity.

3) 6G Base Stations (6GBSs): As critical components in BECS, 6GBSs provide reliable communication services to devices and serve as blockchain maintenance nodes responsible for transaction bookkeeping and block packaging. Each 6GBS consists of three components:

\begin{itemize}
\item{Registration and Certification Component (RCC): The RCC is responsible for managing the identities of devices and issuing and verifying certificates.}
\item{Computation Component (CC): The CC is responsible for maintaining the blockchain.}
\item{Storage Component (SC): The SC is responsible for storing data from devices and the blockchain.}
\end{itemize}



In contrast to the static client-server architecture of 5G MEC, CRP and CRR in BECS do not represent fixed device roles but are instead transient functional states of computing devices, determined by their real-time computational surplus or deficit. This fluidity and flexibility are essential to unlocking the potential of computing resource sharing in 6G CPN. Meanwhile, with the support of blockchain and evolutionary algorithms, computing devices can autonomously match and schedule supply and demand, thereby meeting the ubiquitous and heterogeneous computing demands anticipated in future networks. To ensure this sharing is always productive and aligns with the network's goal of enhancing overall computational capability, the BECS mechanism is founded on the principle that tasks are offloaded to devices with superior capacity. This approach prevents non-productive resource cycling and stabilizes the market, thereby ensuring that sharing is driven by genuine computing resource need and ultimately fostering a trusted ecosystem.

This study primarily focuses on efficient computing resource sharing. Accordingly, the computing device $D_d$ is denoted as $D_d=\{\varsigma_d,\psi_d\}$, where $\varsigma_d$ is usually measured by device's CPU \cite{18}. The inherent heterogeneity of the 6G CPN is captured by the diverse characteristics of these three layers, where devices possess varying computing capacities, service prices, and are subject to different network conditions. Meanwhile, $\mathbb{T} = \{T_1, \ldots, T_t, \ldots, T_T\}$ denotes the set of all computing tasks. The tuple $T_t = \{\varphi_t, \xi_t, \tau_t, d_t\}$ describes the computing task $T_t$. $\varphi_t$ and $\xi_t$ can be obtained using methods described in \cite{19}, such as graph analysis. Notations that will be used are presented in Table \ref{not}.

\begin{table}[h]
	\begin{center}
	\caption{Summary of Notations}
	\label{not}
	\renewcommand{\arraystretch}{1.3}
	\begin{tabular}{lp{5.6cm}}
	\hline
	\textbf{Notation} & \textbf{Definition} \\
	\hline
	$\mathbb{D}$ & the set of all computing devices \\
	$\mathbb{U}$ & the set of user computing devices \\
	$\mathbb{E}$ & the set of edge computing devices \\
	$\mathbb{C}$ & the set of cloud computing devices \\
	$\mathbb{T}$ & the set of all computing tasks \\
	$\varsigma_d$ & the computing capacity (in CPU cycles/s) of $D_d$ \\
	$\psi_d$ & the service price of $D_d$ per unit of time \\
	$\varphi_t$ & the data size of $T_t$ \\
	$\xi_t$ & the required computing amount (in CPU cycles) of $T_t$ \\
	$\xi_d$ & the average computing amount (in CPU cycles) of tasks assigned to $D_d$ \\
	$\tau_t$ & the maximum time consumption allowed of $T_t$ \\
	$d_t \in \{u_t, e_t, c_t\}$ & the final execution device for $T_t$ \\
	$p_u$, $p_n$ & the transmission power of $U_u$ and $U_n$ \\
	$\sigma_0^2$ & the noise power \\
	$B$ & the bandwidth of each subchannel \\
	$N_d=\{U,E,C\}$ & the number of $D_d$ in each layer \\
	$\rho_d$ & the computing resources occupancy of $D_d$ in each layer \\
	$\alpha_d$ & the number of tasks processed per unit of $D_d$'s computing capacity \\
	$p_{d^\ast\to d,t}$ & the probability of offloading $T_t$ from $D_{d^*}$ to $D_d$, and satisfies $\sum_{d \in \{u, e, c\}} p_{d^* \to d,t} = 1 $ \\
	$\kappa_d$ & the effective capacitance coefficient of $D_d$ \\
	$P_{ide}$ & the minimum value of the objective function \\
	\hline
	\end{tabular}
	\end{center}
\end{table}

\subsection{Communication Model}

In BECS, a Non-Orthogonal Multiple Access (NOMA) based communication model is considered for the 6G uplink, which allows multiple user devices to transmit data to the 6GBS over the same frequency band, thereby significantly enhancing spectral efficiency \cite{20,78,79}. Successful decoding in NOMA hinges on the Successive Interference Cancellation (SIC) technique at the receiver \cite{88,59}. This requires the signals to be decoded sequentially, which is predicated on a known channel gain order. Specifically, the receiver decodes signals in descending order of their channel gains; crucially, the power allocation is constrained to guarantee a minimum data rate for weak users, ensuring fairness and satisfying their Quality of Service lower bound.

Without loss of generality, the users are indexed such that their channel gains are in descending order, i.e., $g_1 > \cdots > g_u > \cdots > g_U$. Consequently, when decoding the signal for user $U_u$, signals from users with stronger channel conditions ($U_1, \dots, U_{u-1}$) have already been cancelled. The signals from users with weaker channel conditions ($U_{u+1}, \dots, U_U$) constitute the co-channel interference. The resulting data transmission rate for $U_u$ is therefore calculated via Shannon's theorem as follows:

\begin{equation}
	R_u = B \log_2 \left(1 + \frac{p_u g_u}{\sigma_0^2 + \sum_{n=u+1}^{U} p_n g_n}\right).
\end{equation}

If $U_u$ offloads $T_t$ to $D_d$, the data transmission latency can be expressed as follows:

\begin{equation}
	L_{u,t}^{tra} = \frac{\varphi_t}{R_u}.
\end{equation}

Therefore, the communication energy consumption of $U_u$ can be calculated as follows:

\begin{equation}
	E_{u,t}^{tra} = p_u L_{u,t}^{tra}.
\end{equation}

\subsection{Computing Model}

In the considered three layers computing resource sharing structure, the computing tasks of a CRR can be executed in any device of CRP. However, computing devices vary in terms of task transfer and processing capabilities. This intense competition for computing resources at each layer, inevitable in 6G CPN with numerous CRRs and CRPs, necessitates balancing the task load across the system's layers. Generally, with limited computing resources at each layer, tasks may have to wait for an available processor. Therefore, the M/M/c model \cite{21,81,82}, which describes task offloading as a Poisson process with an average arrival rate of $\lambda_t$, can be used to model task processing delays. Based on the superposition property of the Poisson process and the offloading interactions among the three layers of computing resources, the average task arrival rate for each layer can be calculated as

\begin{equation}
    \lambda_d = \begin{cases}
		\sum_{t=1}^{T} (\lambda_t \times p_{d^* \to u,t}) & \text{if } d = u \\
		\sum_{t=1}^{T} \sum_{d^* \in \{u, e\}} (\lambda_t \times p_{d^* \to e,t}) & \text{if } d = e \\
		\sum_{t=1}^{T} (\sum_{d^* \in \{u, e, c\}} \lambda_t \times p_{d^* \to c,t}) & \text{if } d = c
		\end{cases}.
\end{equation}

Therefore, the average time consumption of each task at each layer, encompassing both the queuing and the execution times, can be calculated as

\begin{equation}
	L_{d,t}^{com} = \frac{C(N_d, \rho_d) \rho_d}{\lambda_d (1 - \rho_d)} + \frac{\xi_t}{\varsigma_d},
\end{equation}
where $\rho_d$ can be calculated as $\rho_d = \frac{\lambda_d}{N_d \alpha_d\varsigma_d}$. $C(N_d, \rho_d)$ is known as Erlang's C formula, which can be calculated as

\begin{equation}
	C(N_d, \rho_d) = \frac{\frac{(N_d \rho_d)^{N_d}}{N_d!}}{\sum_{k=0}^{N_d-1} \frac{(N_d \rho_d)^k}{k!} + \frac{(N_d \rho_d)^{N_d}}{N_d! (1 - \rho_d)}}.
\end{equation}

At the same time, the energy consumption of $D_d$ in each layer during the execution of computing tasks can be calculated as follows \cite{22}:

\begin{equation}
	E_{d,t}^{com} = \kappa_d \xi_t \varsigma_d^2,
\end{equation}
where $\kappa_d$ is depending on the chip architecture.

\subsection{Service Model}

1) Computing occupancy: As a direct indicator of computing resource utilization, it is quantified by the number of devices participating in computing resource sharing while maintaining the provision of all network services. It aims to enhance the overall breadth of resource engagement across the system, which can be defined as follows: 
\begin{equation}
	O_{tot}=\sum_{n=1}^{D}x_{n},
\end{equation}
with
\begin{equation}
	x_{n}=\begin{cases}1 & \text{Device $d_n$ is occupied} \\   0 & \text{Device $d_n$ is free}\end{cases}.
\end{equation}


2) Privacy entropy: Beyond the privacy of task content, the decentralization of computing devices in the 6G CPN introduces significant privacy risks stemming from the offloading patterns themselves. Deterministic offloading strategies, while optimal in terms of performance, generate predictable behavioral patterns. As these decisions are often correlated with the user's wireless channel state—and consequently their location—a curious adversary can analyze offloading metadata (e.g., task size and frequency) to infer sensitive information. To mitigate such statistical inference attacks, privacy entropy is employed to quantify the unpredictability of the task allocation process \cite{23,83,84}. A higher entropy indicates a more disordered allocation pattern, which makes it computationally difficult for an adversary to perform accurate inferences, thereby enhancing user privacy. The types of offloaded data vary according to the computing tasks. Thus, in BECS, we set the relationship between $T_t$ and $p_{d^\ast\to d,t}$ to be a one-to-one correspondence, and the privacy entropy of $T_t$ in $D_d$ can be calculated as

\begin{equation}
    H_{d}=-\sum_{t=1}^{T} p_{d^*\to d,t}\log_2p_{d^*\to d,t}.
\end{equation}

3) Load balancing: As a key metric for evaluating computing device workloads, load balancing aims to equalize each device's load to the average while ensuring an equitable and efficient distribution of workload based on device capabilities. Concentrating multiple tasks on a single device reduces computational efficiency and increases energy consumption. The standard deviation of tasks and the computing capabilities of devices can indicate whether they are load-balanced \cite{24}. Therefore, the load balancing of $D_d$ can be calculated as

\begin{equation}
    B_d=\sqrt{\frac{\sum_{d=1}^D\left(\lambda_d\xi_t-\varsigma_d\right)^2}{D}}.
	\label{11}
\end{equation}

4) Sharing revenue: A suitable revenue can encourage the participation of computing devices in computing resource sharing, thereby enhancing the overall resource utilization. The logarithmic utility function is used to quantify the sharing revenue of $D_d$, which can be calculated as

\begin{equation}
	R_d=\ln \left(1+\beta_1 \psi_d\frac{\xi_t}{\varsigma_d} + \beta_2 \varsigma_d\right).
	\label{12}
\end{equation}

\subsection{Premise Assumptions}

The assumptions and requirements specified below are applied consistently throughout this article unless otherwise stated.

1) The proposed computing resource sharing operates within a metropolitan area, where edge and cloud nodes constitute a well-connected computing infrastructure, linked by a high-capacity optical backbone.

2) Computing tasks can be offloaded from the outer to the inner layer as depicted in Fig. \ref{str}, or within the same layers, provided that the computing capacity of CRP exceeds that of CRR.

3) Each user device is equipped with a single antenna, facilitating real-time communication with the 6GBS.

4) The communication latency on the optical backbone is assumed to be negligible in comparison to the dominant delays introduced by wireless access and task computing \cite{61,89}.

5) Based on the permissioned blockchain, user registration information is visible only to 6GBSs within the blockchain, whereas the status of computing resources is visible to all users.

6) All users securely deliver their keys through a secure channel.

\section{Computing Allocation Based on NSGA-III-KDR}

In this section, we first introduce the proposed computing allocation MOOP. Subsequently, we explain the principle of kernel distance-based Dominance Relation and the optimization process based on NSGA-III-KDR.

\subsection{MOOP of Computing Allocation}


The core of our computing allocation mechanism is formulated as a MOOP. The goal is to intelligently assign computing tasks to available resources in the 6G CPN, balancing multiple, often conflicting, performance objectives. These objectives span from system efficiency and cost to user-centric metrics like privacy and service quality.

To comprehensively address the critical requirements of the 6G CPN ecosystem, including Service Quality and Green Sustainability, System Stability and Efficiency, User Privacy, and Economic Viability, we identify six representative objectives that capture the fundamental trade-offs inherent in decentralized resource sharing. First, the system cost objectives are defined, which include the total time consumption ($L_{tot}$) and total energy consumption ($E_{tot}$). These serve as key metrics for evaluating computing efficiency and energy sustainability in green 6G networks.

\textbf{Total Time Consumption ($L_{tot}$):} As formulated in (\ref{13}), this metric captures the overall latency. It is the sum of the computation time ($L^{com}$) across all three layers and the wireless transmission time ($L^{tra}$) incurred when user devices offload tasks.
 
\begin{equation}
    L_{tot}=\sum_{t=1}^T\sum_{d\in\{u,e,c\}}p_{d^*\to d,t}L_{d,t}^{com}+\sum_{t=1}^T\sum_{u=1}^Up_{u\to d,t}L_{u,t}^{tra},
	\label{13}
\end{equation}

\textbf{Total Energy Consumption ($E_{tot}$):} Similarly, as shown in (\ref{14}), this objective accounts for the total energy spent. It comprises the energy for task execution ($E^{com}$) on the designated devices and the energy for wireless transmission ($E^{tra}$) from user devices.

\begin{equation}
    E_{tot}=\sum_{t=1}^T\sum_{d\in\{u,e,c\}}p_{d^*\to d,t}E_{d,t}^{com}+\sum_{t=1}^T\sum_{u=1}^Up_{u\to d,t}E_{u,t}^{tra}.
	\label{14}
\end{equation}


Next, a set of objectives is defined to address key aspects of system performance, including resource utilization ($O_{ave}$), privacy entropy ($H_{ave}$), load balancing ($B_{tot}$), and CRP revenue ($R_{tot}$).

\textbf{Average Resource Utilization ($O_{ave}$):} As formulated in (\ref{15}), the average resource utilization quantifies the extent of resource engagement by representing the ratio of occupied devices ($O_{tot}$) to the total number of devices ($D$).

\begin{equation}
    O_{ave}=\frac1DO_{tot},
	\label{15}
\end{equation}

\textbf{Average Privacy Entropy ($H_{ave}$):} As shown in (\ref{16}), the average privacy entropy serves as an indicator of user privacy protection. A higher individual entropy ($H_{d,t}$) reflects a more disordered and less predictable task allocation pattern, thereby increasing the difficulty for adversaries to infer sensitive information.

\begin{equation}
    H_{ave}=\frac1D\sum_{d\in\{u,e,c\}}H_{d,t},
	\label{16}
\end{equation}

\textbf{Total Load Balancing ($B_{tot}$):} As expressed in (\ref{17}), the total load balancing metric is computed by aggregating the individual load balance indicators $B_d$ across all devices. The underlying indicator $B_d$, defined in (\ref{11}), corresponds to the standard deviation of workloads, where a smaller value indicates a more balanced load distribution and contributes to the mitigation of computational bottlenecks.

\begin{equation}
    B_{tot}=\sum_{d\in\{u,e,c\}}B_d,
	\label{17}
\end{equation}

\textbf{Total CRP Revenue ($R_{tot}$):} As given in (\ref{18}), the total CRP revenue is obtained by aggregating the individual revenues ($R_{d}$) of all resource-providing devices. The individual revenue function $R_{d}$, detailed in (\ref{12}), is formulated as a logarithmic utility to model economic incentives, capturing diminishing returns and promoting broad participation.

\begin{equation}
    R_{tot}=\sum_{d\in\{u,e,c\}}R_d.
	\label{18}
\end{equation}



Assigning tasks to free computing devices can effectively improve resource utilization. However, focusing solely on utilization improvement is insufficient, as the objectives within the 6G CPN are inherently conflicting. While individual metrics such as time and energy consumption are well-established \cite{2,3,7}, the novelty of this study lies in holistically modeling, for the first time, the complex interplay among performance, efficiency, security, and economy. For instance, minimizing time and energy consumption often leads to workload centralization, which directly conflicts with the goal of equitable load balancing. Similarly, maximizing privacy entropy may necessitate non-optimal routing that degrades performance, whereas a strategy aimed at maximizing provider revenue may create economic tensions that impact overall resource occupancy. This landscape of competing demands necessitates a solution that seeks the best possible compromise. It is worth noting that although other performance indicators (e.g., jitter or packet loss) exist, they are inherently correlated with the selected metrics (e.g., latency) or are less critical under the assumption of a high-capacity optical backbone. Therefore, these six objectives are identified as the minimally sufficient set for characterizing the system trade-offs without introducing redundant dimensions that could unnecessarily hinder algorithmic convergence. Thus, to scientifically balance these competing objectives, the MOOP is formulated as follows:

\begin{eqnarray}
	&& \max\left\{O_{ave}, H_{ave}, R_{tot}\right\}, \\
	&& \min\left\{L_{tot}, E_{tot}, B_{tot}\right\},
\end{eqnarray}
subject to:
\begin{align}
	&\lambda_d \leq \alpha_d\varsigma_d,  \\
	&0<p_u\leq p_u^{\max},  \\
	&L_{d,t}^{com}\leq\tau_t,  \\
	&E_{d,t}^{com}\leq E_d^{max},  \\  
	&\varsigma_u<\varsigma_e<\varsigma_c, \varsigma_{d^*}<\varsigma_d,  \\
	&\lambda_d\xi_d\leq \varsigma_d
\end{align}

Constraint (21) states that the actual task processing rate of the computing device must not exceed its service rate. Constraint (22) restricts the maximum transmit power of the user device. Constraint (23) guarantees that the task execution time does not exceed its deadline. Constraint (24) specifies that the computing energy consumption should not exceed the device's maximum available energy. Constraint (25) differentiates the computing capacities across three layers and specifies task offloading from devices with lower capacity to those with higher capacity. Constraint (26) mandates that the average rate of incoming computing workload to a device must not exceed its processing capacity.

\subsection{Kernel Distance-based Dominance Relation}

NSGA-III, a state-of-the-art evolutionary algorithm, can perform fast global searches with quality guarantees, effectively preventing the MOOP from converging to local optima. Additionally, it addresses high-dimensional optimization problems by maintaining population diversity through uniformly distributed reference points \cite{25}. However, as evidenced by the complexity of the proposed practical six-objective MOOP, even advanced algorithms encounter significant challenges. The performance of Pareto dominance-based algorithms often suffers from severe dimensionality effects caused by dominance resistance when handling MOOPs with more than three objectives \cite{26}. This phenomenon substantially weakens selection pressure, making it difficult for the algorithm to distinguish among a large number of mutually non-dominated solutions, which is an essential limitation when a diverse set of operational strategies is required for the 6G CPN.
 
To overcome this specific bottleneck within the BECS framework, a more powerful dominance relation is required. NSGA-III-KDR is inspired by the concept of a strengthened dominance relation (SDR), which has been demonstrated to be effective in NSGA-II \cite{27}. Given that the core diversity-preserving mechanism of NSGA-III relies on reference points, a distance-based metric is conceptually a natural choice for enhancing its selection process. Accordingly, we propose a kernel distance-based dominance relation (KDR) in NSGA-III-KDR. It builds upon the theoretical foundation of SDR to replace the original Pareto dominance relation in NSGA-III. This enables the algorithm to evaluate solutions not only based on dominance but also in terms of their proximity to the ideal point, thereby ensuring stronger selection pressure for addressing the complex computing sharing in 6G CPN. Specifically, if solution $X_1$ dominates solution $X_2$ in KDR, then (\ref{KDR}) is satisfied.

\begin{equation}
    \begin{cases}d_k(X_1)<d_k(X_2),&\theta_{X_1X_2}\leq\overline{\theta}\\d_k(X_1)\cdot\frac{\theta_{X_1X_2}}{\overline{\theta}}<d_k(X_2),&\theta_{X_1X_2}>\overline{\theta}\end{cases},
	\label{KDR}
\end{equation}
where $\theta_{X_1X_2}$ represents the acute angle between the two candidate solutions $X_1$ and $X_2$, which can be calculated as $\theta_{X_{1}X_{2}}=\operatorname{arccos}\bigl(F(X_{1}),F(X_{2})\bigr)$, and $\overline{\theta}$ denotes the size of the niche to which each candidate solution belongs, and can be set to the $\left\lfloor \frac{|P|}{2} \right\rfloor$-th minimum element of
\begin{equation}
\left\{ \min_{q \in P \setminus \{p\}} \theta_{pq} | p \in P \right\},
\end{equation}
where $\theta_{pq}$ is the acute angle between $p$ and $q$ of any pair of candidate solutions.

The kernel distance is chosen because, as the number of objectives increases, neither the Euclidean distance nor the Mahalanobis distance can accurately reflect the crowding degree between individuals \cite{28}. Therefore, NSGA-III-KDR utilizes the kernel distance from the point $X$ to the ideal point $P_{ide}$ to measure the similarity between them in handling high-dimensional problems \cite{29}. It can be calculated as follows:

\begin{equation}
    d_k(X)=\sqrt{2-2\exp{\left(-\frac{\sum_{i=1}^m(\|f_i(X)-P_{ide}\|^2)}{2\sigma^2}\right)}}.
\end{equation}

\subsection{Encoding of Computing Resources}

Evolutionary algorithms utilize the concept of population evolution to tackle practical MOOPs. Here, individuals in a population are represented by a series of numbers, each mapped to potential solutions of a MOOP through specific encoding. Thus, encoding is crucial for implementing population evolution in practical MOOPs. 

\begin{figure}[h]
	\centering
	\includegraphics[scale=0.4]{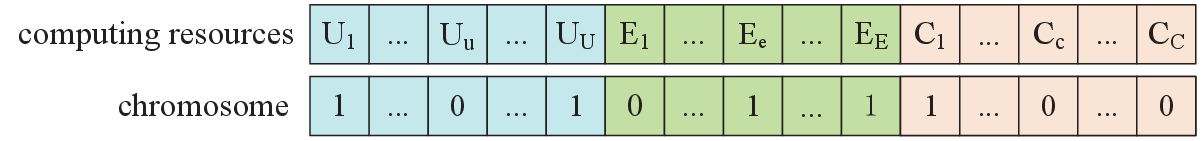}
	\caption{Encoding of computing resources.}
	\label{code}
\end{figure}

In BECS, the occupancy status of computing devices is encoded as genes, while the computing resources involved in sharing are encoded as chromosomes, constituting the entire CRP for evolution. As illustrated in Fig. \ref{code}, a gene value of 0 indicates a free computing device, whereas a value of 1 signifies that this device is occupied. This method allows BECS to integrate various types of computing resources, thereby building a generalized computing resource sharing platform. Since the occupancy status of computing devices directly correlates with gene encoding, and chromosomes relate to computing allocation strategies, dynamic and fine-grained updates of computing resources are enabled.

\subsection{Population Evolution}

To adapt to the dynamic 6G CPN environment, in which task arrivals and resource availability fluctuate over time, BECS operates not as a one-time process but as an adaptive control loop. When a CRR requests computing resources in the network, the computing resource sharing mechanism is triggered. Subsequently, the system executes the NSGA-III-KDR-based computing allocation scheme to match the optimal CRP according to the current network state. NSGA-III-KDR essentially follows the algorithmic framework of NSGA-III, as shown in Algorithm \ref{EVO}. During the evolution process, the chromosomes in the initial population $X^1$ generate entirely new chromosomes (computing allocation strategies) through crossover and mutation operations. The crossover operation enhances chromosome diversity, while the mutation operation, under specific conditions, modifies individual genes to seek those with higher adaptability. In each iteration, the population generated by crossover and mutation merges with the original, then executing a Non-KDR-dominated sort that replaces the non-dominated sort in NSGA-III, thereby effectively enhancing the diversity of the computing resource population. Subsequently, by associating chromosomes with reference points in the hyper-plane and executing the Niche-Preservation Operation, the next generation population can be generated. After $TI$ iterations of NSGA-III-KDR, the final population $X^{TI}$, containing the ultimate computing allocation strategy, is obtained.

\begin{algorithm}[ht]
	\renewcommand{\algorithmicrequire}{\textbf{Input:}}
	\renewcommand{\algorithmicensure}{\textbf{Output:}}
	\caption{Population Evolution}
	\label{EVO}
	\begin{algorithmic}[1]
		\REQUIRE The initialized population $X^1$, the total number of iterations $TI$, and the population size $PS$       
		\ENSURE the final population $X^{TI}$
		\FOR{$j=1$ to $TI$}
		\FOR{the chromosome in $X^j$}
		\STATE Evaluate objective functions by (13), (14), (15), (16), (17) and (18);
		\ENDFOR
		\STATE $X^j$ conducts crossover and mutation operation to produce $S^j$, 
		\STATE Generate a merged population $Y^j = X^j \bigcup S^j$ with $2PS$ population size;
		\FOR{merged population $Y^j$}
		\STATE Execute Non-KDR-dominated sort; 
		\ENDFOR
		\STATE Determine the reference point on the hyper-plane;
		\STATE Associate chromosomes and reference points;
		\STATE Evaluate niche-preservation operation, generate a new population $X^{j+1}$;
		\ENDFOR
		\STATE \textbf{return} $X^{TI}$
	\end{algorithmic}  
\end{algorithm}


\subsection{Optimum Selection}

The final population $X^{TI}$ comprises a set of the best feasible solutions for computing allocation, termed the Pareto optimal set. Each solution in this set represents a different trade-off among the conflicting objectives and is optimal in the sense that no objective can be improved without degrading at least one other. However, in practical computing allocation problems, the CRR needs to match only a specific CRP, necessitating a decision-making process to select a single strategy for implementation. Therefore, the Technique for Order Preference by Similarity to Ideal Solution (TOPSIS) \cite{85}, in conjunction with entropy weighting, is utilized to select the most suitable solution for a specific requirement from the set of Pareto optimal solutions. As a method approximating the ideal solution, TOPSIS rapidly identifies the optimal solution by ranking all solutions according to their distance from the positive (negative) ideal solution. Additionally, entropy weighting helps eliminate the arbitrariness of subjectively determined weights. This combination enables an objective and fair determination of the final computing allocation strategy from the $X^{TI}$. The specific algorithm is detailed in \cite{30}.

By executing NSGA-III-KDR in BECS, CRP can set $\psi_d$ more reasonably by analyzing the optimal solution. Simultaneously, CRR can select the most appropriate CRP for task offloading based on the optimal solution. Consequently, both parties engaged in computing resource sharing achieve reliable, dynamic, and secure computing allocation and trading.

\section{Secure Computing Trading with Privacy Protection}

In this section, we focus on the principles of the proposed pseudonym scheme and secure computing trading under pseudonyms in BECS.

\subsection{System Initialization}

In this phase, upon inputting the security parameter $\lambda$, the system administrator selects a secure elliptic curve $\mathcal{E}:y^2 = x^3 + ax + b \mod p$. The group $\mathcal{G}$ is an additive elliptic curve group with order $q$ and generator $g$ defined over $\mathcal{E}$, where $p$ and $q$ are two large prime numbers. Subsequently, 6GBS $B_b$ chooses a random number $b \in Z_q^*$ as its private key and calculates the public key as $bg$. Additionally, it selects a collision-resistant hash function $\mathcal{H()}:\{0,1\}^* \to \{0,1\}^l$ and makes $\{\mathcal{E}, \mathcal{G}, a, b, p, q, \mathcal{H()}\}$ publicly available in blockchain.

\subsection{Computing Devices Registration}

All devices must be registered upon entering the system. $D_d$ first selects $d \in Z_q^*$ as the private key and then calculates the public key as $dg$. Subsequently, it sends its real identity $ID_d$ and the public key $dg$ to the nearest $B_b$ via a secure channel. After receiving the message from $D_d$, $B_b$ stores $\mathcal{H}(\mathcal{H}(ID_d) \parallel dg)$ in SC, while simultaneously generating and sending $D_d$'s identity certificate: $Cert_d = Sig_b(\mathcal{H}(ID_d) \parallel dg)$.

\subsection{Pseudonym Generation}

When $D_d$ participates in computing resource sharing, it must first send its public key $dg$ and identity certificate $Cert_d$ to the nearest $B_b$. After $B_b$ verifies the legitimacy of $D_d$'s identity, it generates a pseudonym for $D_d$, as detailed in Algorithm \ref{pyg}. Based on the Schnorr protocol \cite{31}, $D_d$ proves possession of the private key $d$ to $B_b$ without revealing it, by utilizing the public parameter $g$ and random numbers. $B_b$ verifies the legitimacy of $D_d$'s identity by confirming the correctness of $D_d$'s proof and response. Utilizing non-interactive zero-knowledge protocol, $D_d$ acquires a pseudonym $(x,y)$ for communication from $B_b$ while preserving privacy.

\begin{algorithm}[ht]
	\renewcommand{\algorithmicrequire}{\textbf{Input:}}
	\renewcommand{\algorithmicensure}{\textbf{Output:}}
	\caption{Pseudonym Generation}
	\label{pyg}
	\begin{algorithmic}[1]
		\REQUIRE $D_d$'s private key $d$, public key $dg$.
		\ENSURE $D_d$'s pseudonym $(x, y)$.
		\STATE $D_d \longrightarrow B_b$: Sends $(\tilde{x}=g, \tilde{y}=dg)$.
		\STATE $B_b \longrightarrow D_d$: Chooses $\gamma \in Z_q^*$, sends $x = \gamma \tilde{x}$.
		\STATE $D_d \longrightarrow B_b$: Calculates  $y = dx$, then chooses $\delta \in Z_q^*$, and calculates commitment $K=\delta x$, challenge $\epsilon = \mathcal{H}(x \parallel y \parallel K)$, response $M=\delta+\epsilon d \mathrm{~mod~} q$, sends $y, K$, and $M$.
		\STATE $B_b$: Calculates $\epsilon' = \mathcal{H}(x \parallel y \parallel K)$ to validate $Mx \stackrel{?}{=} K+\epsilon' y$. 
		\IF{verification succeeds} 
		\STATE $B_b$: Stores $D_d$'s pseudonym $(x, y)$ in SC.
		\STATE $D_d$: Stores its pseudonym $(x, y)$, and the transcript $T=(K,M)$.
		\ENDIF
		\STATE \textbf{return} $(x, y)$
	\end{algorithmic}  
\end{algorithm}

The pseudonym $(x,y)=(\gamma g, \gamma dg)$, constructed using $D_d$'s public key and random numbers, cannot establish its legitimacy and thus requires $B_b$ to issue the corresponding certificate, as shown in Algorithm \ref{cri}. Using a non-interactive zero-knowledge protocol, $B_b$ generates a commitment and a response that include its private key. $D_d$ then verifies these to confirm that $B_b$, who possesses the private key $b$, generated the certificate. Upon successful verification, $D_d$ stores the certificate.

\begin{algorithm}
	\renewcommand{\algorithmicrequire}{\textbf{Input:}}
	\renewcommand{\algorithmicensure}{\textbf{Output:}}
	\caption{Certificate Issuance}
	\label{cri}
	\begin{algorithmic}[1]
	\REQUIRE $D_d$'s pseudonym $(x, y)$; $B_b$'s private key $b$, public key $h = bg$.
	\ENSURE $D_d$'s pseudonym certificate $\varpi$.
	\STATE $B_b \longrightarrow D_d$: Chooses $\phi \in Z_q^*$, and calculates commitment $O=\phi g$, challenge $\zeta = \mathcal{H}(x \parallel y \parallel O)$, response $P=\phi+\zeta b \mathrm{~mod~} q$, sends $O$ and $P$.
	\STATE $D_d$: Calculates $\zeta' = \mathcal{H}(x \parallel y \parallel O)$ to verify $P g \stackrel{?}{=} O + \zeta' h$, if verification succeeds, then store the certificate $\varpi=(O,P)$.
	\STATE \textbf{return} $\varpi$
	\end{algorithmic}
\end{algorithm}

\subsection{Pseudonymous Computing Trading}

The pseudonymous computing trading in BECS is structured as a hybrid on-chain/off-chain protocol designed to ensure both real-time responsiveness and verifiable trust. This architecture decouples rapid off-chain execution from asynchronous on-chain settlement, thereby mitigating the inherent latency caused by blockchain consensus. While a CRP initially registers its identity and core capabilities on-chain to establish a foundational trust anchor, its dynamic operational status is maintained off-chain. When a CRR ($D_1$) initiates a request, the NSGA-III-KDR matches it with an optimal CRP ($D_2$). The algorithm operates on a near real-time view of resource availability, maintained by the 6GBS and continuously updated based on both the canonical on-chain state and transient off-chain ``soft-state" messages. The trading process consists of two primary phases:

1) Off-Chain Execution: First, both parties verify their pseudonyms and certificates according to Algorithm \ref{idv} to establish mutual trust through a secure off-chain channel. Subsequently, they negotiate and finalize an off-chain agreement by cryptographically signing the terms of service. Importantly, upon reaching this agreement, $D_2$ immediately broadcasts a signed lightweight status message to the nearest 6GBS. This message serves as an off-chain ``soft-state" update, temporarily marking the resource as ``in use." This rapid propagation ensures that $D_2$ is instantly excluded from the pool of available resources in any concurrent NSGA-III-KDR matching process, thereby preventing double booking. To ensure system resilience and prevent resource deadlocks, this soft state is linked to a predefined timeout, and the resource is automatically released if an on-chain settlement transaction is not detected within this period. This mechanism allows the CRR to offload computing tasks for immediate execution without waiting for blockchain confirmation.

2) On-chain Settlement: Second, this phase begins only after the CRP completes the task and delivers the results to the CRR. Upon successful verification of the outcome, a formal transaction is generated. This transaction includes the essential details of the trade, such as digital signatures from both parties, hashed references to the off-chain agreement, and proof of task completion. The transaction is subsequently submitted to the blockchain. The successful inclusion of this transaction in a block finalizes the trade, generating an immutable and auditable record while formally updating the canonical state of the CRP's resources on the ledger.

\subsection{Block Generation and Pseudonymous Update}


Following the transaction submission, the 6GBSs serve as trusted consensus nodes responsible for validating transactions and maintaining the ledger. The selection of the 6GBS responsible for block generation is determined by the Proof of Trust and Adjustment (PoTA) consensus mechanism, a lightweight and efficient protocol that we introduced in our previous work \cite{17}. Specifically, the PoTA protocol selects a bookkeeping node based on a dynamically computed score. This score comprehensively evaluates both a node's global trust, derived from historical interaction ratings, and its relevant resource service attributes. This mechanism maintains a balance between security and efficiency, making it a practical and effective consensus solution for dynamic 6G CPN environments. The 6GBS with the highest score is authorized to aggregate all relevant information, such as transaction details, digital signatures, and resource status updates, into a new block. This block is then broadcast for validation and subsequently appended to the blockchain.

Subsequently, the CRP's computing resource status is formally updated and finalized within the new block, serving as the canonical on-chain record of the state change that was already broadcast off-chain as a soft state. This two-step process ensures that the NSGA-III-KDR algorithm operates on the most up-to-date information from off-chain soft-state messages, while the blockchain maintains the ultimate ground truth for consistency and security. Before re-engaging in computing resource sharing, the CRP and CRR must request new pseudonyms and certificates from the nearest 6GBS to ensure their identities remain unlinkable. Similarly, the 6GBS verifies the legitimacy of the CRP and CRR identities using Algorithm \ref{idv} and subsequently issues them new pseudonyms and certificates.

\begin{algorithm}[h]
	\renewcommand{\algorithmicrequire}{\textbf{Input:}}
	\renewcommand{\algorithmicensure}{\textbf{Output:}}
	\caption{Identity Verification}
	\label{idv}
	\begin{algorithmic}[1]
	\REQUIRE $D_1$'s pseudonym $(x_1, y_1)$, transcript $T_1=(K_1,M_1)$, and certificates $\varpi_1=(O_1,P_1)$ are issued by $B_1$; $B_1$'s public keys $h_1=b_1g$.
	\ENSURE $D_2$ accepts or rejects $D_1$'s identity.
	\STATE $D_1 \longrightarrow D_2$: Sends transcript $T_1=(K_1,M_1)$ and certificates $\varpi_1=(O_1,P_1)$ under the pseudonym $(x_1, y_1)$.
	\STATE $D_2$: Calculates $\epsilon_1' = \mathcal{H}(x_1 \parallel y_1 \parallel K_1)$ to verify the authenticity of pseudonym: $M_1 x_1 \stackrel{?}{=} K_1 + \epsilon_1' y_1$.
	\\ Then Calculates $\zeta_1' = \mathcal{H}(x_1 \parallel y_1 \parallel O_1)$ to verify the legitimacy of pseudonym and certificate: $P_1 g \stackrel{?}{=} O_1 + \zeta_1' h_1$.
	\STATE If all verifications are successful, $D_2$ accepts $D_1$'s identity. Else, $D_2$ rejects $D_1$'s identity.
	\STATE \textbf{return} $D_2$ accepts or rejects $D_1$'s identity.
	\end{algorithmic}
\end{algorithm}


\section{Security Analysis and Discussion}

In this section, we analyze the security of BECS and discuss the computational complexity of NSGA-III-KDR.

\subsection{Threat Model and Adversary Assumptions}

To facilitate formal analysis, a comprehensive threat model is adopted, encompassing two primary types of adversaries:

\begin{itemize}
    \item \textbf{External Adversary ($\mathcal{A}$):} An entity that is not a legitimate participant in the 6G CPN. Its capabilities are limited to eavesdropping on communication channels, with the objective of compromising user privacy via traffic analysis or forging messages to obtain unauthorized access.
    
    \item \textbf{Malicious Internal Node ($\mathcal{B}$):} A registered participant within the BECS system (e.g., a compromised CRP or CRR) may comply with protocol specifications but still attempt to exploit its privileges to deanonymize transaction partners, repudiate actions, or disrupt system operations by injecting false information.

	\item \textbf{Statistical Inference Adversary ($\mathcal{C}$):} A curious-but-honest internal participant (e.g., an MEC server) that correctly processes tasks while passively analyzing historical offloading metadata \cite{64}. Its goal is to compromise user privacy by inferring wireless channel states from observed offloading patterns, thereby revealing the user's physical location and behavioral characteristics.
\end{itemize}

The security analysis relies on standard cryptographic assumptions, including the computational hardness of the discrete logarithm problem and the collision resistance of the hash function $\mathcal{H}$.

\subsection{System Security}

The permissioned blockchain architecture establishes a foundational layer of defense against both external and internal adversaries.

\textbf{1) Access Control and Authentication:} This mechanism constitutes the first line of defense against the external adversary $\mathcal{A}$. By mandating that all devices register through the RCC and obtain verifiable credentials, the system effectively prevents unauthorized entities from joining the network, thereby mitigating risks such as impersonation and unauthorized access to resources.

\textbf{2) Integrity and Non-repudiation:} Blockchain immutability serves as a direct countermeasure to threats posed by the malicious internal node $\mathcal{B}$. It ensures that malicious participants cannot tamper with transaction records to alter agreed-upon terms or repudiate their actions. All activities are irreversibly recorded, providing a verifiable audit trail that enforces accountability.

\textbf{3) Resilience and Availability:} The blockchain's distributed architecture is inherently resilient to targeted attacks by a powerful adversary $\mathcal{B}$ capable of compromising one or several 6GBSs. By eliminating single points of failure, the system ensures service continuity and preserves data integrity even in the presence of partial compromise.

\textbf{4) Resistance to Statistical Inference Attacks:} The privacy entropy maximization objective in the proposed MOOP provides a statistical defense against inference attacks launched by the malicious internal node $\mathcal{B}$. By promoting diverse and unpredictable task offloading behaviors, this mechanism obfuscates user behavioral patterns, thereby complementing the direct identity unlinkability provided by the pseudonym scheme.

\textbf{5) Balancing Transparency and Privacy:} BECS resolves the inherent tension between market transparency and user privacy by architecturally decoupling the public visibility of resource status from the identities of participants. Global transparency of resource availability is an essential requirement for any decentralized resource marketplace, as it enables requesters to efficiently discover potential providers. BECS ensures that such transparency does not compromise user privacy by incorporating a robust pseudonym scheme, where all transactions are performed through temporary, unlinkable identifiers. Consequently, it prevents both external and internal adversaries from constructing long-term behavioral profiles or linking a user's activities across multiple transactions, thereby achieving an effective balance between market efficiency and strong privacy preservation.

\subsection{Pseudonym and Certificate Security}

The pseudonym scheme is equipped with specific cryptographic properties to mitigate the capabilities of both adversaries, $\mathcal{A}$ and $\mathcal{B}$, across a range of attack scenarios.

\textbf{1) Uniqueness and Authenticity:} This property mitigates impersonation attacks launched by both adversaries, $\mathcal{A}$ and $\mathcal{B}$. The zero-knowledge protocol-based generation process ensures that only the legitimate owner of a private key $d$ can generate a valid pseudonym and its corresponding proof. As a result, it becomes computationally infeasible for an adversary to forge a valid pseudonym for another user, thereby ensuring that each identity within a transaction is authentic.

\textbf{2) Unforgeability:} Certificate unforgeability directly prevents both adversaries from obtaining illegitimate trust within the system. Since generating a valid certificate requires access to the private key $b$ of a 6GBS, no adversary can fabricate a fraudulent certificate, thereby preventing impersonation of certified and trustworthy nodes.

\textbf{3) Unlinkability:} This property plays a critical role in defending against the privacy-compromising objectives of both adversary types.
\begin{itemize}
    \item Against the external adversary $\mathcal{A}$, the use of fresh, random nonces $\gamma$ in each session ensures that pseudonyms appear as independent and uncorrelated values on the public channel. This renders traffic analysis and activity inference through eavesdropping ineffective.
    \item Against the malicious internal node $\mathcal{B}$, the zero-knowledge-based interactions and session-specific pseudonyms prevent it from linking its current transaction partner to any previous or future transactions. This prevents malicious peers from constructing a user's long-term behavioral profile.
\end{itemize}

\textbf{4) Traceability:} This feature is specifically designed as a deterrent and mitigation strategy against the malicious internal node $\mathcal{B}$. While unlinkability safeguards users from peer inference, authorized traceability ensures that when malicious behavior is verified, the 6GBSs can revoke the anonymity of the offending node and expel it from the system. This accountability mechanism deters internal misbehavior without compromising user anonymity.

\subsection{Privacy-Performance Trade-off}

The inclusion of privacy entropy as a simultaneous optimization objective introduces a fundamental trade-off with performance-oriented metrics. A strategy optimized exclusively for performance, such as minimizing latency or energy consumption, would inevitably favor deterministic offloading decisions and thus produce predictable, low-entropy patterns. As established in the threat model, these predictable patterns constitute the primary attack surface exploited by $\mathcal{C}$ to compromise user privacy. Maximizing privacy entropy necessitates introducing a degree of non-determinism into the allocation process to obscure these patterns, thereby obfuscating the correlation between offloading decisions and the user's state.

This inherent conflict is not a limitation to be eliminated but rather a fundamental challenge that must be intelligently managed, validating the necessity of a MOOP. The NSGA-III-KDR algorithm is designed not to find a single utopian solution but to identify the Pareto-optimal front, which consists of a set of non-dominated solutions, each representing a quantitatively optimal and practically meaningful trade-off between robust privacy and high performance. The final policy is then selected using TOPSIS to align with specific application requirements, thereby transforming the trade-off into a mechanism for flexible and context-aware optimization.

\subsection{Computational Complexity of NSGA-III-KDR}

In BECS, NSGA-III-KDR uses Non-KDR-dominated sort instead of Non-Dominated Sort from NSGA-III to enhance solution diversity, while maintaining the same computational complexity. Namely, with $L$ optimization objectives and a population size of $X$, NSGA-III-KDR exhibits a computational complexity of $\mathcal{O}(LX^2)$. Specifically, calculating the kernel distance for each solution and determining the angle between any two solutions each incur a complexity of $\mathcal{O}(L)$. Each solution's distance to the ideal point is calculated individually, contributing to a complexity of $\mathcal{O}(LX)$. For dominance judgment, each pair of solutions undergoes one angle calculation and comparison, leading to an overall complexity of $\mathcal{O}(LX^2)$. Thus, the computational complexity of each iteration can be approximated as $\mathcal{O}(LX^2)$. Compared to the NSGA-III-based scheme \cite{6}, NSGA-III-KDR offers more diverse solutions, improving the probability of superior computing allocation strategies.

\section{Simulation Results and Analysis}

\begin{table}[h]
	\begin{center}
	\caption{Key Parameters}
	\label{para}
	\renewcommand{\arraystretch}{1.3}
	\begin{tabular}{| >{\centering\arraybackslash}m{3cm} | >{\centering\arraybackslash}m{4.9cm} |}
	\hline
	\textbf{Parameters} & \textbf{Values} \\
	\hline
	Main physical machine & Intel i7-12700@2.1GHz with 32GB RAM \\
	\hline
	Operating systems & Windows 11 \& Ubuntu 24 \\
	\hline
	Number of devices & $\mathbb{U}$: 300; $\mathbb{E}$: 200; $\mathbb{C}$: 100 \\
	\hline
	Transmit power & 20$\sim$30 dBm \\
	\hline
	Noise power & -97 dBm \\
	\hline
	Channel gain & 2.15 dBi \\
	\hline
	Bandwidth & 20 MHz \\
	\hline
	Average arrival rate & 100 tasks/s \\
	\hline
	Computing Capacity \cite{62,38} & $\mathbb{U}$: 0.6$\sim$10 TFLOPS; $\mathbb{E}$: 10$\sim$1000 TFLOPS; $\mathbb{C}$: $>$1000 TFLOPS\\
	\hline
	Effective capacitance coefficient \cite{39} & $10^{-29}$ \\
	\hline
	Data size & 500$\sim$3000 KB \\
	\hline
	CPU cycles per byte & 1000 cycles/byte \\
	\hline
	Cryptographic libraries & PBC and Openssl \\
	\hline
	Probability of offloading & $\mathbb{U}$: 0.5, $\mathbb{E}$: 0.3, $\mathbb{C}$: 0.2 \\
	\hline
	Service price & $\mathbb{U}$: 0.1, $\mathbb{E}$: 1, $\mathbb{C}$: 2 \\
	\hline
	$\beta_1$, $\beta_2$; $\sigma$ & 0.6, 0.4; 1\\
	\hline
	\end{tabular}
	\end{center}
\end{table}

In this section, we initially compare the performance of the proposed NSGA-III-KDR with NSGA-III and MOEA/D applied in computing allocation, as well as the NSGA-II-SDR that inspired us. Subsequently, we test the performance of four algorithms in optimizing the proposed computing allocation MOOP. Finally, we simulate the performance of the proposed pseudonym scheme. The configurations of critical parameters are detailed in Table \ref{para}. The code will be available at https://github.com/QuinYim/BECS.

\subsection{Simulation of Proposed NSGA-III-KDR}

In this part, we compare the proposed NSGA-III-KDR with state-of-the-art evolutionary algorithms NSGA-III \cite{6}, MOEA/D \cite{7}, and NSGA-II-SDR \cite{27}, utilizing the PlatEMO platform \cite{40}. The widely used SDTLZ \cite{25}, MaF \cite{42}, and SMOP \cite{43} test suites are employed as the benchmarks. Meanwhile, IGD \cite{44} and PD \cite{45} are selected as performance evaluation metrics. IGD comprehensively measures the convergence and diversity of the algorithms, where a smaller IGD value indicates better performance. PD primarily reflects the diversity of the algorithms, with a larger PD value indicating greater population diversity. The crossover probability is set to 1, the mutation probability is set to 1/D, and their distribution indicator is set to 20, where D represents the length of the decision variable. All algorithms are executed for 1000 iterations over 30 independent times on different test problems, and the average values are taken. The performance of four algorithms is compared under different numbers of objectives and benchmarks, as shown in Table \ref{igdpd}, where “$+$”, “$-$”, and “$=$” indicate that the result is significantly better, significantly worse, and statistically similar to that obtained by NSGA-III-KDR, respectively.

\begin{table*}[ht]
	\centering
	\caption{IGD and PD value of NSGA-III, MOEA/D, NSGA-II-SDR, and NSGA-III-KDR on SDTLZ1, SDTLZ2, MaF1, MaF2, SMOP1, and SMOP2 with 5, 10, and 15 objectives. The best result in each row is highlighted.}
	\label{igdpd}
	\renewcommand{\arraystretch}{1.3}
	\scriptsize
	  \begin{tabular}{|c|c|c|c|c|c|c|c|c|c|}
	  \hline
	  \multirow{2}{*}{Problem} & \multirow{2}{*}{M} & \multicolumn{4}{c|}{IGD}      & \multicolumn{4}{c|}{PD} \\
  \cline{3-10}          &       & NSGA-III & MOEA/D & NSGA-II-SDR & NSGA-III-KDR & NSGA-III & MOEA/D & NSGA-II-SDR & NSGA-III-KDR \\
	  \hline
	  \multirow{3}{*}{SDTLZ1} & 5     & 4.0919e-1 = & 1.1456e+0 - & 7.2486e-1 - & \textbf{4.072e-1} & 1.0645e+7 - & 5.7924e+6 - & 1.8399e+7 - & \textbf{4.7052e+7} \\
  \cline{2-10}          & 10    & \textbf{1.7833e+1 +} & 3.1719e+1 + & 3.7045e+1 + & 1.3076e+2 & \textbf{2.3247e+10 +} & 4.3704e+9 + & 7.6415e+8 + & 1.0865e+10 \\
  \cline{2-10}          & 15    & \textbf{6.1421e+2 +} & 8.2078e+2 + & 8.4100e+2 + & 6.4468e+3 & \textbf{2.8108e+12 +} & 4.7472e+11 + & 9.5939e+7 - & 5.2570e+9 \\
	  \hline
	  \multirow{3}{*}{SDTLZ2} & 5     & 1.1871e+0 = & 3.1693e+0 - & 4.3480e+0 - & \textbf{1.1838e+0} & 1.9000e+7 - & 1.1551e+7 - & 1.6919e+4 - & \textbf{9.6971e+7} \\
  \cline{2-10}          & 10    & 6.8214e+1 - & 1.1354e+2 - & 1.4962e+2 - & \textbf{6.3516e+1} & 6.3070e+10 - & 8.2536e+9 - & 6.7541e+8 - & \textbf{1.0334e+10} \\
  \cline{2-10}          & 15    & 2.3123e+3 - & 3.6367e+3 - & 3.6612e+3 - & \textbf{2.1131e+3} & 1.3576e+13 - & 4.7141e+11 - & 8.7418e+8 - & \textbf{2.9897e+11} \\
	  \hline
	  \multirow{3}{*}{MaF1} & 5     & 2.2118e-1 - & 1.6710e-1 - & \textbf{1.4031e-1 +} & 1.6200e-1 & 1.7519e+7 - & 3.8550e+6 - & 1.9648e+7 - & \textbf{2.4256e+7} \\
  \cline{2-10}          & 10    & 3.1732e-1 - & 3.9154e-1 - & \textbf{2.9201e-1 +} & 3.1425e-1 & 8.9830e+9 - & 1.5818e+9 - & 6.1301e+9 - & \textbf{1.2171e+10} \\
  \cline{2-10}          & 15    & \textbf{3.8109e-1 +} & 4.7345e-1 - & 4.2790e-1 - & 4.0302e-1 & 1.9765e+11 = & 5.7436e+10 - & 5.9691e+10 - & \textbf{2.0030e+11} \\
	  \hline
	  \multirow{3}{*}{MaF2} & 5     & 1.4175e-1 - & 1.5433e-1 - & 1.3740e-1 - & \textbf{1.2311e-1} & 1.8852e+7 - & 1.3030e+7 - & 1.6372e+7 - & \textbf{2.1218e+7} \\
  \cline{2-10}          & 10    & 2.7622e-1 - & 3.0954e-1 - & 4.0882e-1 - & \textbf{2.5215e-1} & 8.5156e+9 - & 3.8250e+9 - & 8.3977e+9 - & \textbf{1.3398e+10} \\
  \cline{2-10}          & 15    & \textbf{3.2353e-1 +} & 3.8517e-1 + & 5.6897e-1 - & 4.6420e-1 & \textbf{2.5275e+11 +} & 1.1452e+11 - & 2.4382e+10 - & 1.3115e+11 \\
	  \hline
	  \multirow{3}{*}{SMOP1} & 5     & 1.5928e-1 - & 2.2883e-1 - & 3.4716e-1 - & \textbf{1.3947e-1} & 3.6432e+6 - & 3.7787e+6 - & 6.8850e+6 - & \textbf{2.5642e+7} \\
  \cline{2-10}          & 10    & 4.2414e-1 - & 3.9985e-1 - & 4.9668e-1 - & \textbf{2.8703e-1} & 2.0785e+9 - & 9.4496e+8 - & 1.9355e+9 - & \textbf{1.1620e+10} \\
  \cline{2-10}          & 15    & 4.0587e-1 + & 4.5602e-1 + & \textbf{4.0515e-1 +} & 5.2449e-1 & 8.5548e+10 - & 5.5518e+9 - & 7.0498e+10 - & \textbf{1.0424e+11} \\
	  \hline
	  \multirow{3}{*}{SMOP2} & 5     & 3.8732e-1 - & 5.8495e-1 - & 5.4909e-1 - & \textbf{3.6202e-1} & 6.2126e+6 - & 6.4898e+6 - & 1.1547e+7 - & \textbf{3.9017e+7} \\
  \cline{2-10}          & 10    & 9.1827e-1 - & 5.8728e-1 - & 8.6089e-1 - & \textbf{5.5237e-1} & 3.7243e+9 - & 1.8461e+9 - & 3.3883e+9 - & \textbf{2.1840e+10} \\
  \cline{2-10}          & 15    & \textbf{6.2944e-1 +} & 7.2789e-1 + & 6.8003e-1 + & 9.1024e-1 & 1.7008e+11 = & 1.1430e+10 - & 1.3620e+11 - & \textbf{2.0211e+11} \\
	  \hline
	  \multicolumn{2}{|c|}{+/-/=} & 6/10/2 & 5/13/0 & 6/12/0 &       & 3/13/2 & 2/16/0 & 1/17/0 &  \\
	  \hline
	  \end{tabular}
  \end{table*}
  
It can be concluded from the experimental results that NSGA-III-KDR has the strongest competitiveness, achieving the best IGD and PD numbers of 10 and 15, respectively. This demonstrates that NSGA-III-KDR achieves better performance compared to the other two algorithms, especially in terms of population diversity, offering a richer set of solutions for computing allocation.

\subsection{Simulation of Proposed Computing Allocation Scheme}

In this part, we compare NSGA-III, MOEA/D, NSGA-II-SDR, and NSGA-III-KDR in optimizing the proposed computing allocation MOOP. Specifically, each chromosome in the evolutionary algorithm represents a candidate computing resource allocation scheme for the entire network, where each gene corresponds to the binary occupancy status of a specific computing device. Consequently, the population constitutes a diverse set of potential allocation schemes. To simulate a representative snapshot of the dynamic CPN environment, we initialize the population with 50\% of the computing devices randomly occupied, then perform the optimization using each of the four algorithms separately. This simulation evaluates the algorithm's ability to obtain a high-quality allocation solution within a single decision epoch, demonstrating its effectiveness in adapting to the heterogeneous and resource-constrained conditions of that specific moment. We calculate the changes in computing resources between the final and initial populations, while ensuring algorithmic convergence. Due to the uncertainty of the evolutionary process, we evaluate the performance of the four algorithms using three statistical indicators. Specifically, the overall distribution is presented via box plots (showing the maximum, upper quartile, median, lower quartile, and minimum), the optimal values based on the TOPSIS method are denoted by triangles, and the average values are represented by squares. This multi-dimensional comparison enables a more comprehensive assessment of algorithm performance.

\begin{figure*}[ht]
	\setlength{\abovecaptionskip}{-5pt}
	\setlength{\belowcaptionskip}{-10pt}
	\centering
	\begin{minipage}[t]{0.32\linewidth}
		\centering
		\includegraphics[width=2.56in]{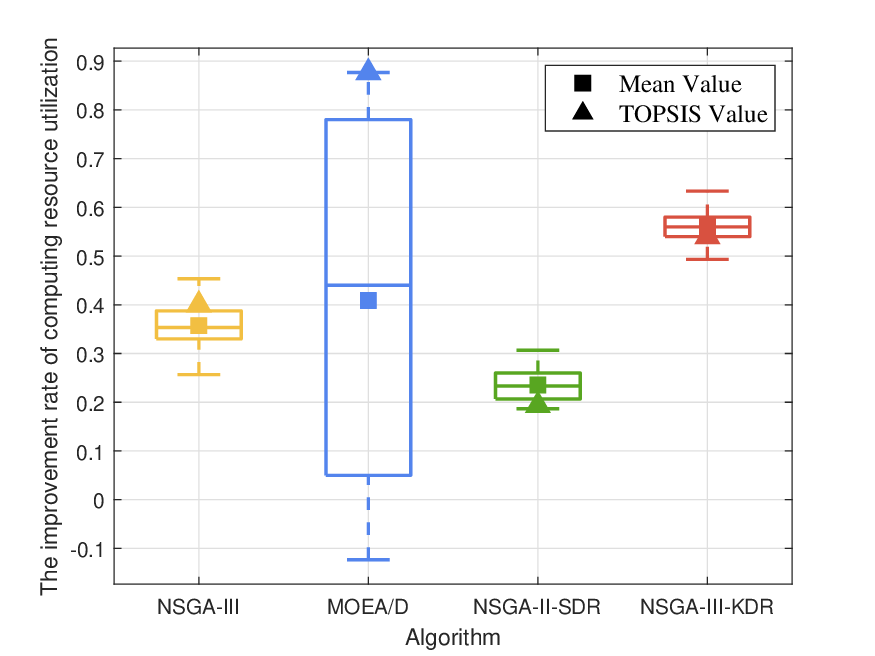}
		\caption{The IR of computing resource utilization.}
		\label{ircr}
	\end{minipage}
	\begin{minipage}[t]{0.32\linewidth}
		\centering
		\includegraphics[width=2.56in]{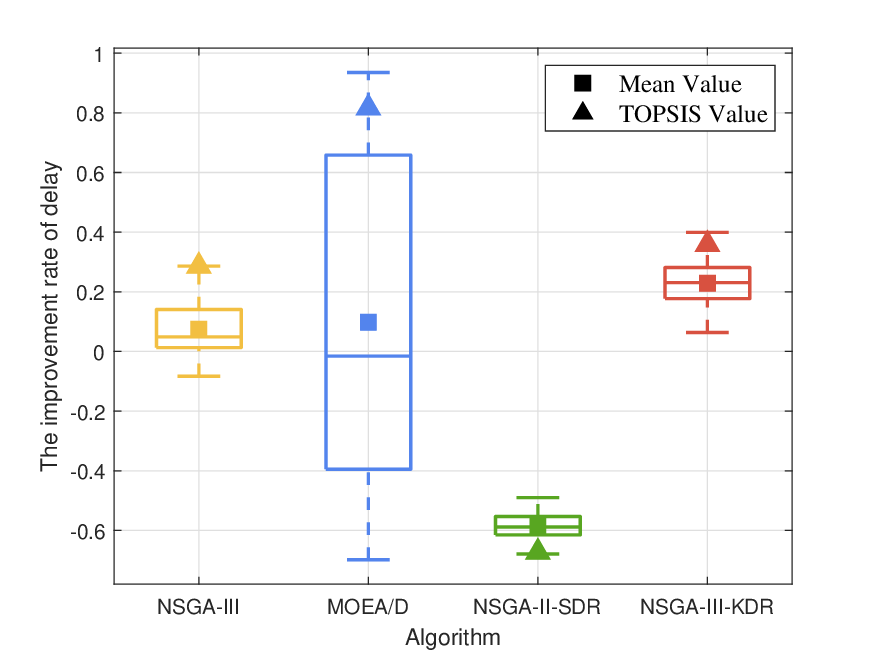}
		\caption{The IR of time consumption.}
		\label{irtc}
	\end{minipage}
	\begin{minipage}[t]{0.32\linewidth}
		\centering
		\includegraphics[width=2.56in]{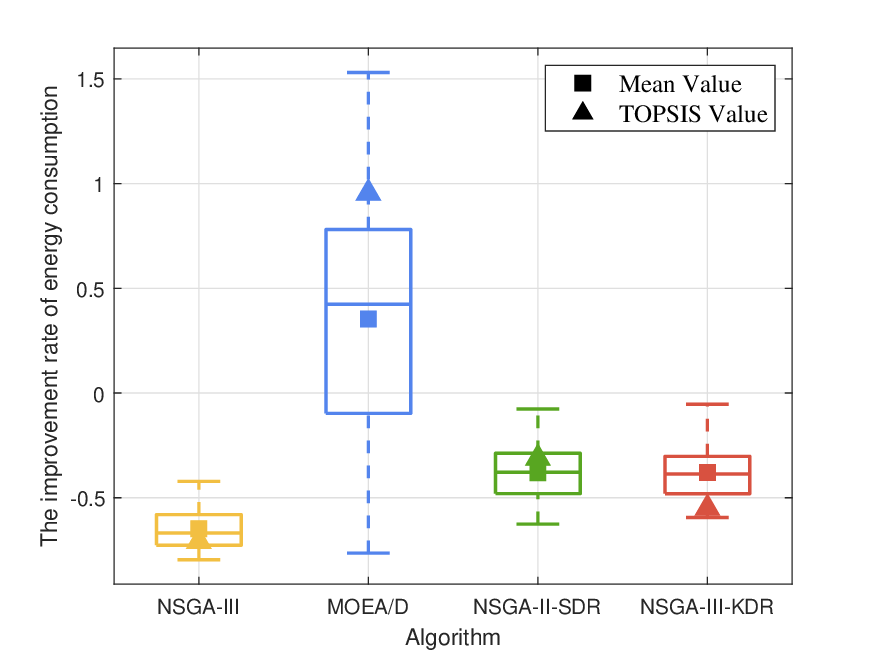}
		\caption{The IR of energy consumption.}
		\label{irec}
	\end{minipage}
	\begin{minipage}[t]{0.32\linewidth}
		\centering
		\includegraphics[width=2.56in]{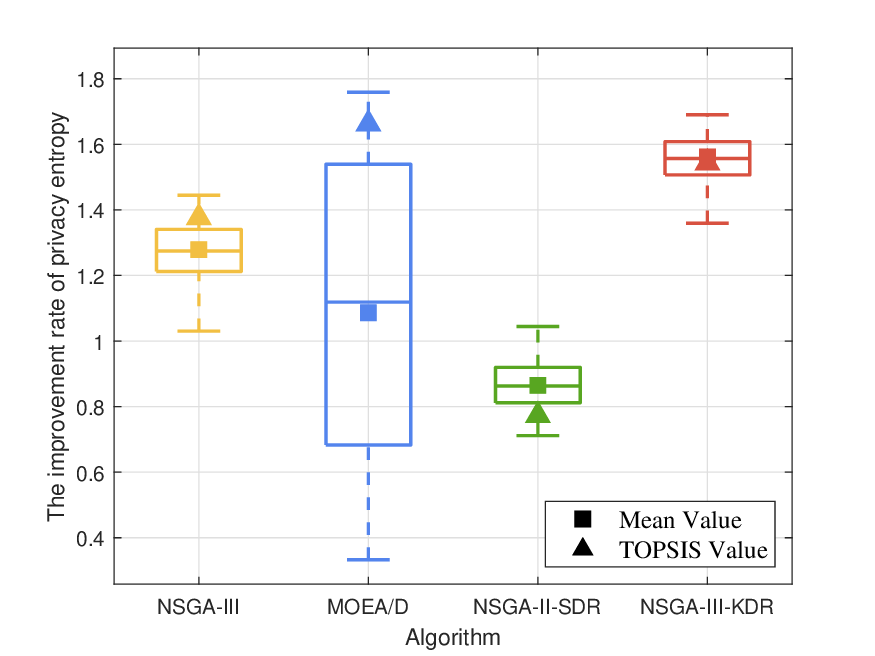}
		\caption{The IR of privacy entropy.}
		\label{irpe}
	\end{minipage}
	\begin{minipage}[t]{0.32\linewidth}
		\centering
		\includegraphics[width=2.56in]{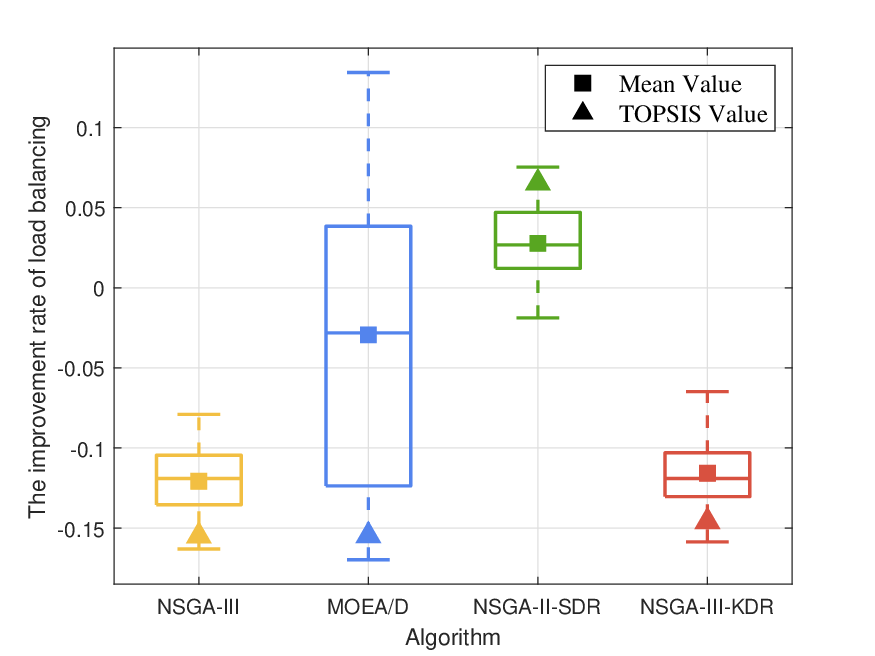}
		\caption{The IR of load balancing.}
		\label{irlb}
	\end{minipage}
	\begin{minipage}[t]{0.32\linewidth}
		\centering
		\includegraphics[width=2.56in]{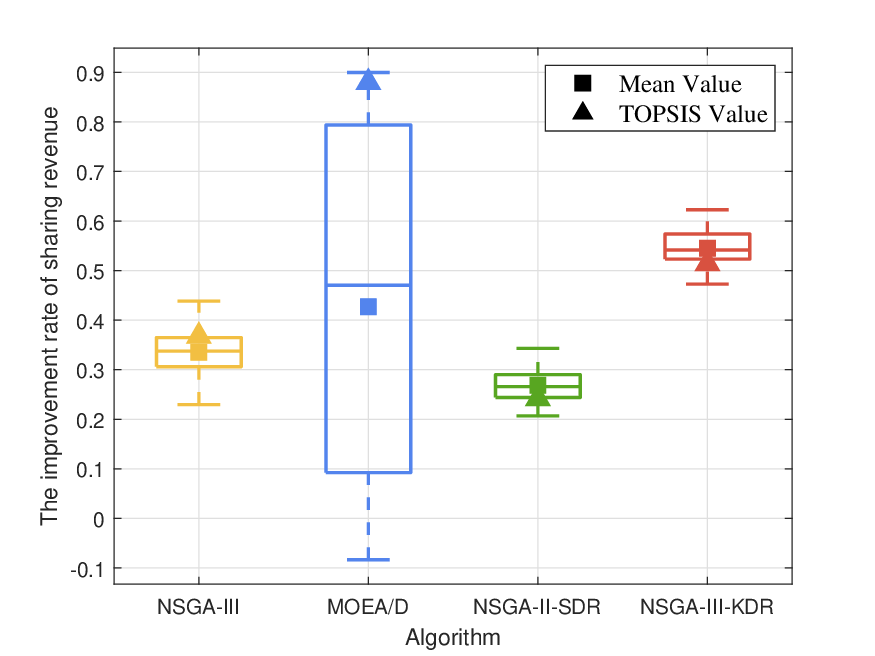}
		\caption{The IR of sharing revenue.}
		\label{irsr}
	\end{minipage}
\end{figure*}

The improvement rate (IR) of computing resource utilization under four algorithms is shown in Fig. \ref{ircr}. Overall, except for MOEA/D, the other three algorithms demonstrate positive optimization in computing resource utilization. NSGA-III-KDR shows the most significant overall and average improvements, indicating superior performance and the most substantial enhancement in resource utilization. NSGA-III and NSGA-II-SDR follow, showing notable but lesser improvements. Although MOEA/D achieves the best optimal solution, its overall performance is suboptimal, characterized by some negatively optimized computing allocation strategies. This issue likely stems from MOEA/D’s predefined fixed neighborhood structures, which may restrict its global search capability and lead to local optimality.

As resource utilization improved, Fig. \ref{irtc} to Fig. \ref{irsr} illustrate the changes in the other five objectives considered by the computing allocation MOOP. Except for load balancing, improvements in all other objectives are observed with the optimization based on NSGA-III-KDR, attributed to enhanced resource utilization. The superior diversity of NSGA-III-KDR facilitates a more even distribution of computing resources, as particularly evidenced by the significant increase in privacy entropy. Although NSGA-III exhibits similar trends, it is less effective than NSGA-III-KDR. Notably, with NSGA-III-KDR and NSGA-III, increased resource utilization results in higher delays and sharing revenue. Meanwhile, energy consumption decreases as more tasks are offloaded to $\mathbb{U}$, leveraging all available resources within 6G CPN comprehensively. The optimization strategy of NSGA-II-SDR favors offloading tasks to $\mathbb{E}$ and $\mathbb{C}$, significantly reducing delay, but uniquely resulting in positive growth in load balancing among the four algorithms. The optimal solution of MOEA/D surpasses all other algorithms, achieving the highest resource utilization and substantial improvements in all objectives, except for load balancing, and the decline in load balancing suggests a rational allocation of computing resources. However, the overall data of MOEA/D is the worst, with a negative optimization of 12.99\%. Furthermore, although its median or average increase in resource utilization is less than that of NSGA-III-KDR, the more substantial improvement in load balancing suggests a less favorable overall evolution. Since the selection of the optimal solution using TOPSIS is random, not every iteration of MOEA/D yields a solution that outperforms those of other algorithms.

\subsection{Simulation of Proposed Pseudonymous Scheme}

\begin{table*}[ht]
	\begin{center}
	\caption{Comparison of computational overhead}
	\label{cco}
	\renewcommand{\arraystretch}{1.3}
	\begin{tabular}{| >{\centering\arraybackslash}m{2cm} | >{\centering\arraybackslash}m{3.5cm} | >{\centering\arraybackslash}m{3.5cm} | >{\centering\arraybackslash}m{3.5cm} | >{\centering\arraybackslash}m{2cm} |}
	\hline
	\textbf{Scheme} & \textbf{SIPG} & \textbf{CIMS} & \textbf{IDMV} & \textbf{Total}\\
	\hline
	Bagga \emph{et al.}'s scheme \cite{32} & $4T_{pm}^{ecc}+2T_{mtp}+1T_{ha}=12.867$ $ms$ & $3T_{pm}^{ecc}+3T_{pa}^{ecc}+3T_{ha}=1.86$ $ms$ & $3T_{bp}+4T_{pm}^{ecc}+4T_{pa}^{ecc}+3T_{ha}=6.847$ $ms$ & $21.574$ $ms$ \\
	\hline
	Shen \emph{et al.}'s scheme \cite{33} & $3T_{bp}+4T_{pm}^{bp}+1T_{pa}^{bp}+2T_{ep}+4T_{ha}=9.505$ $ms$ & $1T_{bp}+1T_{pm}^{bp}+1T_{ep}+1T_{ha}=2.745$ $ms$ & $2T_{bp}+2T_{pm}^{bp}+1T_{pa}^{bp}+1T_{ep}+1T_{ha}=5.481$ $ms$ & $17.731$ $ms$ \\
	\hline
	Yang \emph{et al.}'s scheme \cite{34} & $17T_{pm}^{ecc}+5T_{pa}^{ecc}+11T_{ha}=10.522$ $ms$ & $1T_{ha}=0.001$ $ms$ & $3T_{pm}^{ecc}+4T_{pa}^{ecc}+2T_{ha}=1.86$ $ms$ & $12.383$ $ms$ \\
	\hline
	Wang \emph{et al.}'s scheme \cite{35} & $4T_{pm}^{ecc}+1T_{pa}^{ecc}+2T_{ha}=2.475$ $ms$ & $9T_{pm}^{ecc}+4T_{pa}^{ecc}+5T_{ha}=5.571$ $ms$ & $3T_{pm}^{ecc}+2T_{pa}^{ecc}+4T_{ha}=1.86$ $ms$ & $9.906$ $ms$ \\
	\hline
	Our proposed scheme & $7T_{pm}^{ecc}+1T_{pa}^{ecc}+4T_{ha}=4.331$ $ms$ & $3T_{pm}^{ecc}+1T_{pa}^{ecc}+2T_{ha}=1.857$ $ms$ & $4T_{pm}^{ecc}+2T_{pa}^{ecc}+2T_{ha}=2.476$ $ms$ & $8.664$ $ms$ \\
	\hline
	\end{tabular}
	\end{center}
\end{table*}

In this part, we evaluate the computational overhead of the proposed pseudonym scheme by comparing it with several existing pseudonymous schemes. Each operation and phase is tested 20 times separately, and the average value is recorded as the experimental data to eliminate the influence of hardware fluctuations during the operation. First, the BN curve \cite{46} with a 128-bit security level is selected to implement the bilinear group. The execution time of basic cryptographic operations is shown in Table \ref{etco}. Generally, the execution times required for bitwise XOR and modular multiplication are significantly lower compared to other cryptographic operations, and are thus disregarded. Moreover, we utilize SHA-256 as the hash function. Then, we assess the time costs associated with system initialization and pseudonym generation (SIPG), certificate issuance or message signing(CIMS), and identity or message verification (IDMV) in the pseudonym authentication process, comparing the proposed scheme with those in \cite{32,33,34,35}, as shown in Table \ref{cco}.

The proposed scheme, which is based on the Schnorr protocol, primarily involves operations $T_{pm}^{ecc}$, $T_{pa}^{ecc}$, and $T_{ha}$ in the pseudonym authentication process. The total computational overload is $14T_{pm}^{ecc} + 4T_{pa}^{ecc} + 8T_{ha} = 8.664\ ms$. Conversely, Bagga \emph{et al.}'s scheme \cite{32} uses $T_{mtp}$ and $T_{bp}$ in the SIPG and IDMV, leading to a higher computational overload. Similarly, Shen \emph{et al.}'s scheme \cite{33}, based on bilinear pairing, also incurs a higher computational overload. Although based on ECC, Yang \emph{et al.}'s scheme \cite{34} includes 17 times $T_{pm}^{ecc}$ within SIPG, leading to a significant computational overload. Wang \emph{et al.}'s scheme \cite{35}, similar to the proposed scheme, provides the optimal computational overload in both SIPG and IDMV. However, a total of 16 times $T_{pm}^{ecc}$ yields a marginally greater computational overload than ours. Therefore, our scheme achieves the lowest computational overload compared to other related schemes.

\begin{table}[h]
	\begin{center}
	\caption{Execution time of cryptographic operations}
	\label{etco}
	\renewcommand{\arraystretch}{1.3}
	\begin{tabular}{| c | c | c |}
	\hline
	\textbf{Notation} & \textbf{Operation} & \textbf{Time (ms)} \\
	\hline
	$T_{pm}^{ecc}$ & point multiplication in ECC & 0.618 \\
	\hline
	$T_{pa}^{ecc}$ & point addition in ECC & 0.001 \\
	\hline
	$T_{ha}$ & general hash & 0.001 \\
	\hline
	$T_{ep}$ & exponentiation & 0.011 \\
	\hline
	$T_{bp}$ & bilinear pairing & 1.456 \\
	\hline
	$T_{pm}^{bp}$ & point multiplication in bilinear pairing & 1.277 \\
	\hline
	$T_{pa}^{bp}$ & point addition in bilinear pairing & 0.003 \\
	\hline
	$T_{mtp}$ & hash-to-point in bilinear pairing & 5.197 \\
	\hline
	\end{tabular}
	\end{center}
\end{table}

\begin{figure}[ht]
	\centering
	\includegraphics[scale=0.56]{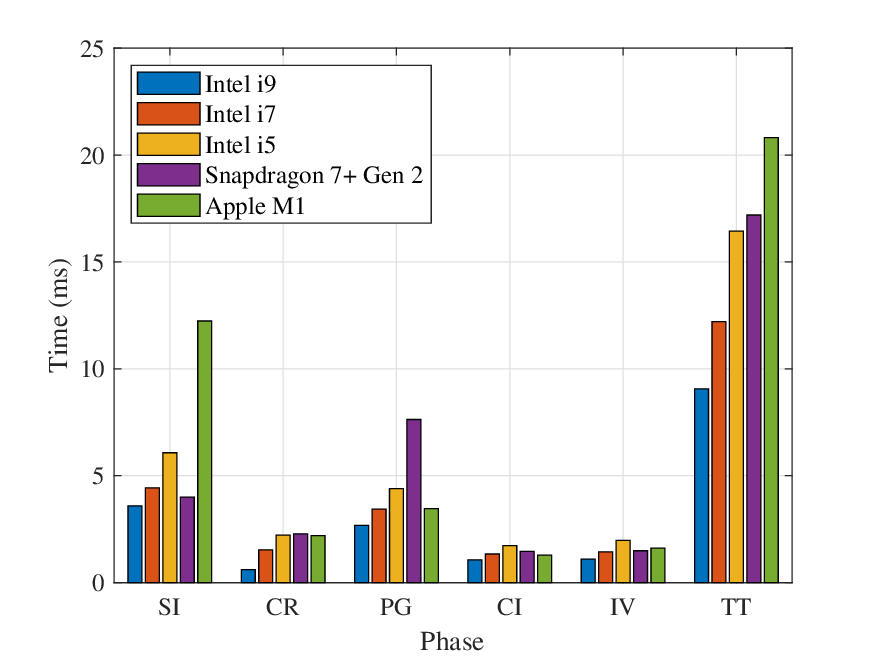}
	\caption{The time consumption on different devices.}
	\label{pt}
\end{figure}

Next, considering the diversity of computational devices, we test the portability of the proposed pseudonym scheme. We consider five different devices, including a workstation with an Intel i9-12900k@3.9GHz and 64GB RAM (Intel i9), a computer with an i7-12700@2.1GHz and 32GB RAM (Intel i7), a computer with an i5-8500@3GHz and 8GB RAM (Intel i5), a smartphone with Snapdragon 7+ Gen 2 and 16GB RAM (Snapdragon 7+ Gen 2), and a MacBook with an M1 chip and 16GB RAM (Apple M1). We test the time consumption for each of the five phases of the pseudonym authentication process on these devices: System Initialization (SI), Computing Devices Registration (CR), Pseudonym Generation (PG), Certificate Issuance (CI), and Identity Verification (IV), including the Total Time (TT) for the entire process, as shown in Fig. \ref{pt}. Time consumption varies across devices due to differences in CPU performance. Specifically, the MacBook requires the longest total time to complete a pseudonym authentication, at $20.813\ ms$. The SI is the most time-consuming, as it involves generating system keys and registering user information. However, typically, SI is only performed once, whereas CR, PG, CI, and IV take less than $8\ ms$ across all devices, with PG taking the longest at $7.628\ ms$ on the smartphone. This demonstrates the good portability and lightweight of the proposed pseudonym scheme.

\section{Conclusion}

This article investigates improving computing resource utilization within 6G CPN and proposes BECS, a privacy-preserving computing resource sharing mechanism. We consider various communication, computing, and service factors in 6G CPN, modeling them as a six-objective MOOP. Meanwhile, we utilize the proposed NSGA-III-KDR to find optimal solutions for this MOOP. Additionally, we introduce a novel pseudonym scheme to protect the privacy of users engaged in computing resource sharing. Extensive simulations demonstrate the effectiveness of BECS. Moving forward, we intend to further explore computing sharing in dynamic real-time scenarios and investigate advanced access control solutions to protect the task payload.

\bibliographystyle{IEEEtran}
\bibliography{references.bib}



\begin{IEEEbiography}[{\includegraphics[width=1in,height=1.25in,clip,keepaspectratio]{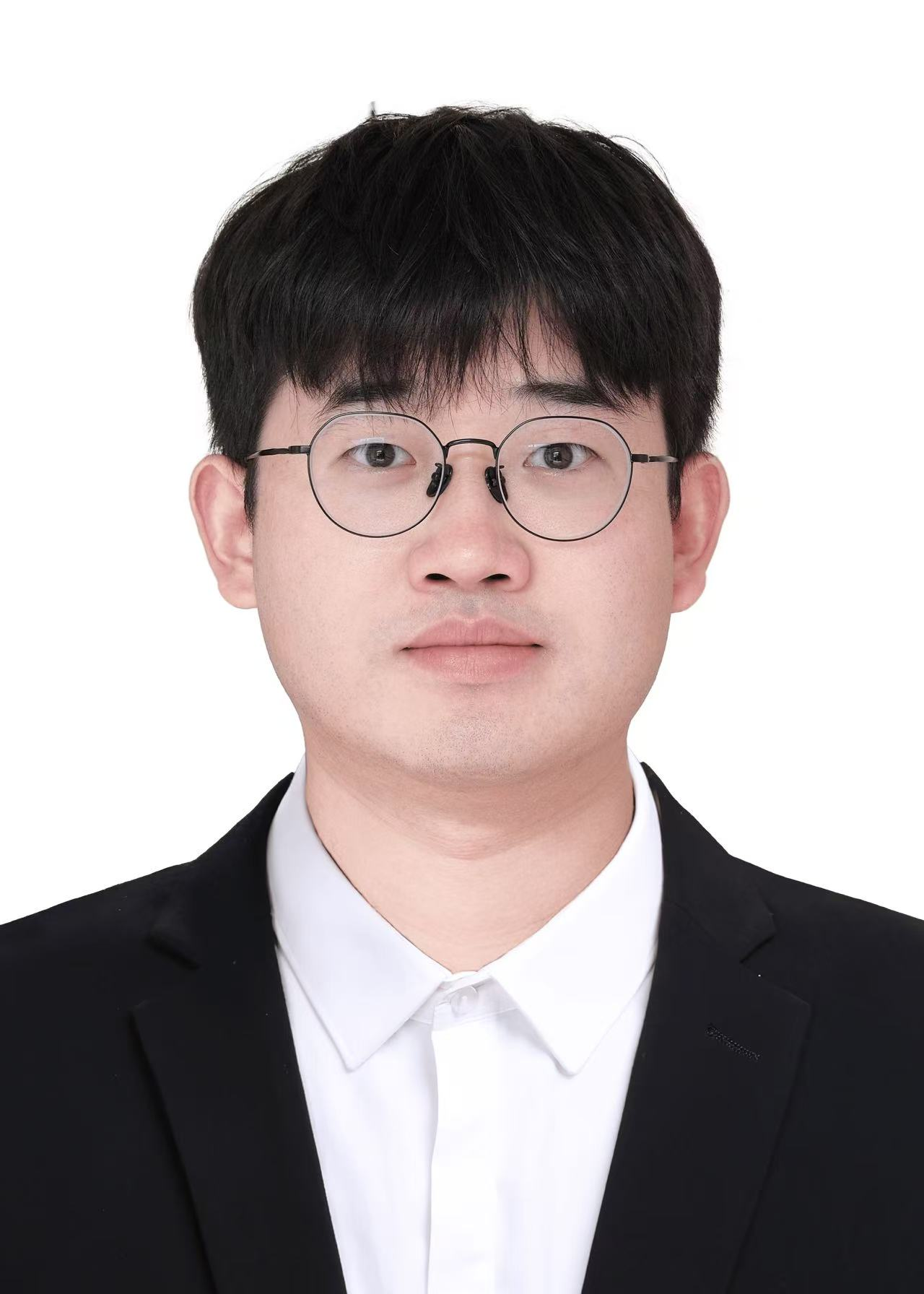}}]{Kun Yan}(Member, IEEE) received the B.S. degree in internet of things engineering from Xi'an University of Technology, Xi'an, China, in 2019, and the Ph.D. degree in cryptography from Xidian University, Xi'an, China, in 2025. He is currently a Postdoctoral Fellow at The Hong Kong Polytechnic University, Hong Kong. His research interests include 6G networks, resource management, internet of things, and blockchain.
\end{IEEEbiography}

\vspace{-33pt}

\begin{IEEEbiography}[{\includegraphics[width=1in,height=1.25in,clip,keepaspectratio]{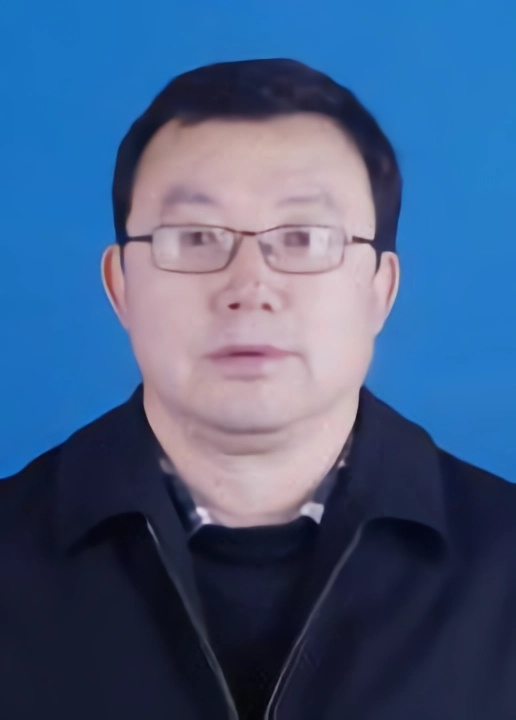}}]{Wenping Ma}(Member, IEEE) received the B.S. and M.S. degrees in fundamental mathematics from Shaanxi Normal University, Xi'an, China, in 1987 and 1990, respectively, and the Ph.D. degree in communication and information system from Xidian University, Xi'an, in 1999, where he is currently a Full Professor with the School of Telecommunications Engineering. His current research interests include information theory, communication theory, blockchain, and 6G networks security.
\end{IEEEbiography}

\vspace{-33pt}

\begin{IEEEbiography}[{\includegraphics[width=1in,height=1.25in,clip,keepaspectratio]{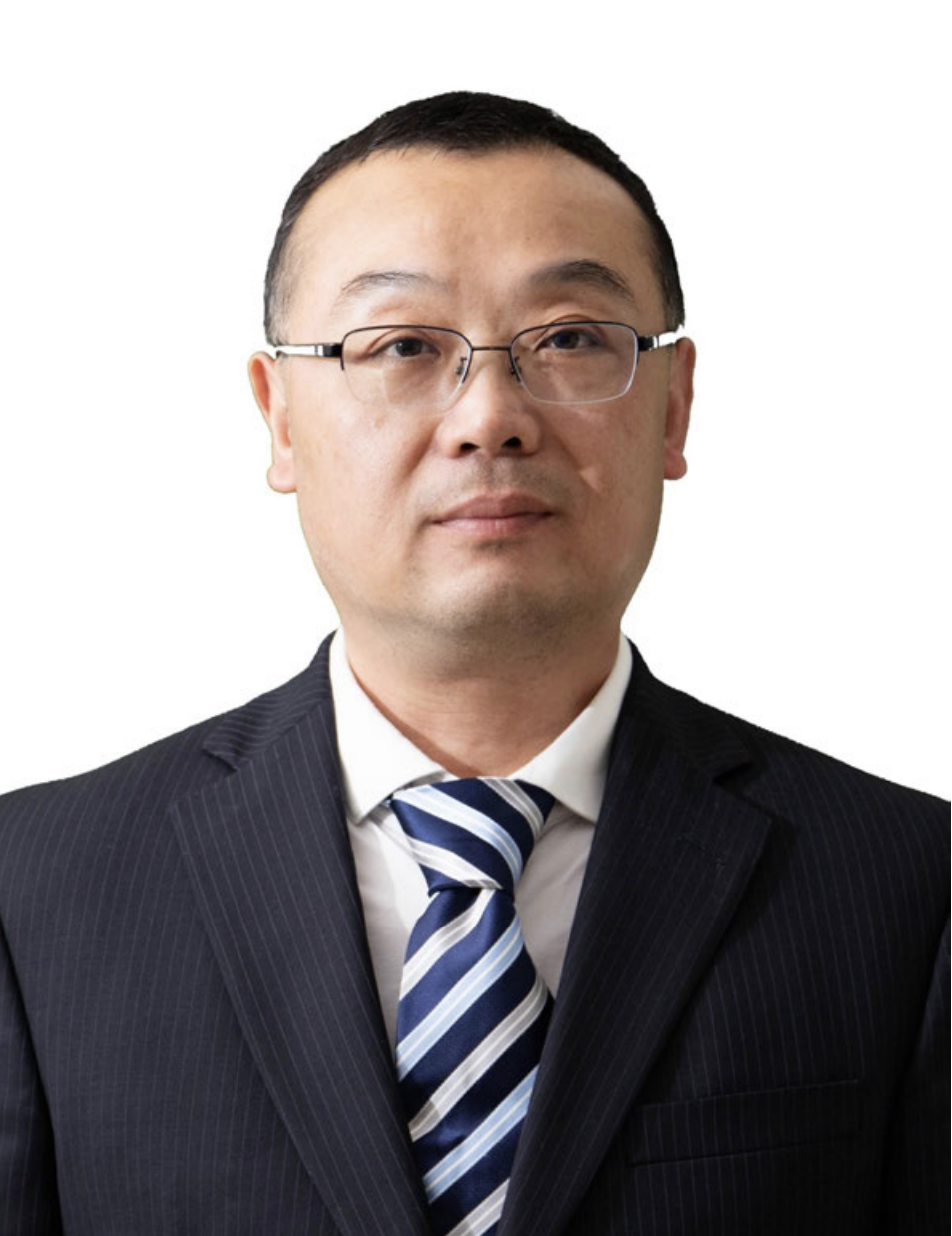}}]{Shaohui Sun}(Senior Member, IEEE) received his Ph.D. from Xidian University, Xi'an, China, in 2003. From March 2003 to June 2006, he was a postdoctoral research fellow at the Datang Telecom Technology and Industry Group, Beijing, China. From June 2006 to December 2010, he worked at the Datang Mobile Communications Equipment Co., Ltd., Beijing. Since January 2011, he has been the Chief Technical Officer with Datang Wireless Mobile Innovation Center of the Datang Telecom Technology and Industry group. His current research interest includes advanced technologies related to B5G/6G.
\end{IEEEbiography}

\end{document}